# EnVision
## Understanding why our most Earth-like neighbour is so different

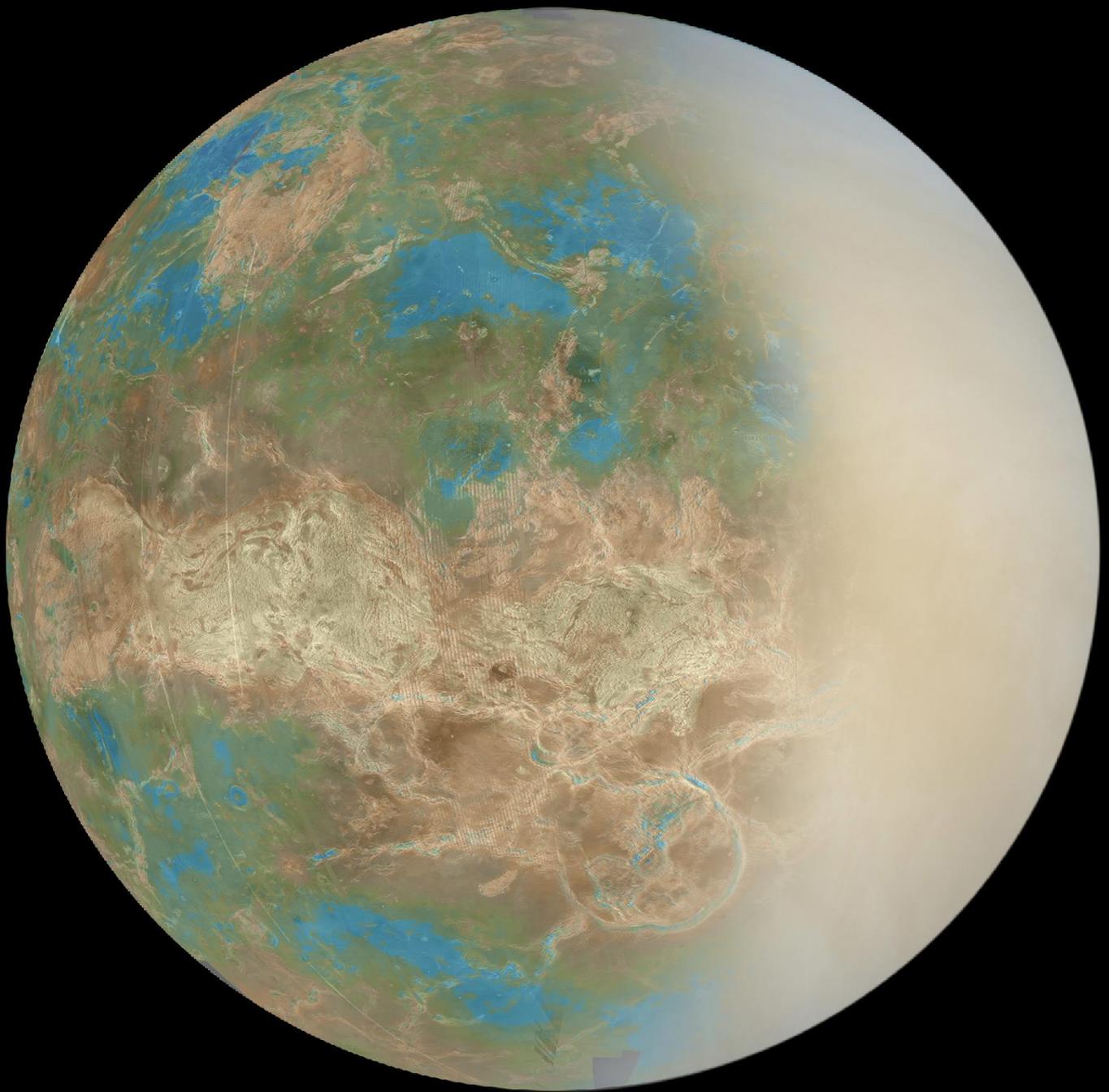

esa — Cosmic Vision 2015-2025 — Science Programme European Space Agency

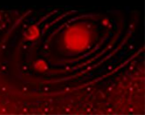

*Proposal Name: EnVision*

*Lead Proposer: Richard Ghail*

*Core Team members*

**Richard Ghail**
*Radar Systems Engineering*
Civil and Environmental Engineering,
Imperial College London, United Kingdom

**Jörn Helbert**
*Thermal Infrared Mapping*
Institute for Planetary Research,
DLR, Germany

**Lorenzo Bruzzone**
*Subsurface Sounding*
Remote Sensing Laboratory,
University of Trento, Italy

**Thomas Widemann**
*Ultraviolet, Visible and Infrared Spectroscopy*
LESIA, Observatoire de Paris,
France

**Philippa Mason**
*Surface Processes*
Earth Science and Engineering,
Imperial College London, United Kingdom

**Colin Wilson**
*Atmospheric Science*
Atmospheric Physics,
University of Oxford, United Kingdom

**Caroline Dumoulin**
*Interior Dynamics*
Laboratoire de Planétologie et Géodynamique de Nantes,
France

**Ann Carine Vandaele**
*Spectroscopy and Solar Occultation*
Belgian Institute for Space Aeronomy,
Belgium

**Pascal Rosenblatt**
*Spin Dynamics*
Royal Observatory of Belgium
Brussels, Belgium

**Emmanuel Marcq**
*Volcanic Gas Retrievals*
LATMOS, Université de Versailles Saint-Quentin, France

**Robbie Herrick**
*StereoSAR*
Geophysical Institute,
University of Alaska, Fairbanks, United States

**Louis-Jerome Burtz**
*Outreach and Systems Engineering*
ISAE-Supaero
Toulouse, France

*Executive Summary*

Why are the terrestrial planets so different? Venus should be the most Earth-like of all our planetary neighbours: its size, bulk composition and distance from the Sun are very similar to those of Earth. Its original atmosphere was probably similar to that of early Earth, with abundant water that would have been liquid under the young sun's fainter output. Even today, with its global cloud cover, the surface of Venus receives less solar energy than does Earth, so why did a moderate climate ensue here but a catastrophic runaway greenhouse on Venus? How and why did it all go wrong for Venus? What lessons can be learned about the life story of terrestrial planets in general, in this era of discovery of Earth-like exoplanets? Were the radically different evolutionary paths of Earth and Venus driven solely by distance from the Sun, or do internal dynamics, geological activity, volcanic outgassing and weathering also play an important part?

ESA's Venus Express a landmark in Venus exploration, answered many questions about our nearest planetary neighbour and established European leadership in Venus research. Focussed on atmospheric research, Venus Express nonetheless discovered tantalising hints of current volcanic activity including a tenfold changes in mesospheric sulphur dioxide, anomalously dark lava surrounding volcanoes, and surface temperature changes that all point towards activity which had not been expected from NASA's Magellan mission of the early 1990s. That mission showed that Venus has abundant volcanic and tectonic features but did not have the resolution or technology necessary to detect geological activity.

We therefore propose EnVision, a medium class mission to determine the nature and current state of geological activity on Venus, and its relationship with the atmosphere, to understand how Venus and Earth could have evolved so differently. EnVision will use a world-leading European phased array synthetic aperture radar, VenSAR, to:

- Obtain images at a range of spatial resolutions from 30 m regional coverage to 1 m images of selected areas; an improvement of two orders of magnitude on Magellan images;
- Measure topography at 15 m resolution vertical and 60 m spatially from stereo and InSAR data;
- Detect cm-scale change through differential InSAR, to characterise volcanic and tectonic activity, and estimate rates of weathering and surface alteration; and
- Characterise of surface mechanical properties and weathering through multi-polar radar data.

Its RIME-heritage subsurface radar sounder, SRS, will:

- Characterise the vertical structure and stratigraphy of geological units including volcanic flows;
- Determine the depths of weathering and aeolian deposits; and
- Discover as yet unknown structures buried below the surface.

VEM, an IR mapper and an IR and UV spectrometer suite, will:

- Search for temporal variations in surface temperatures and tropospheric concentrations of volcanically emitted gases, indicative of volcanic eruptions; and
- Study surface-atmosphere interactions and weathering by mapping surface emissivity and tropospheric gas abundances.

EnVision will also take advantage of its low circular orbit to:

- Provide gravity and geoid data at a geologically-meaningful scale, and
- Measure the spin rate and spin axis variations, to constrain interior structure.

VenSAR, the S-band phased array antenna, has heritage from NovaSAR and Sentinel-1 and is funded by the UK Space Agency and ESA. SRS, derived from RIME on board JUICE, has heritage from MARSIS and SHARAD and is funded by the Italian Space Agency. The Venus Emissivity Mapper, VEM, operating in the infrared, builds on the success of VIRTIS and VMC on Venus Express and has heritage from SOIR, NOMAD, and MERTIS. It is funded by a consortium of the German, French, and Belgian Space Agencies.

The proposed baseline mission is ESA-only, with science payloads funded by ESA member states as outlined above; no hardware contribution from international partners is required. Envision is launched on an Ariane 6.2 with a nominal launch date of 24 October 2029. Following a brief 5-month cruise, the spacecraft will perform a Venus Orbit Insertion manoeuvre using conventional propulsion to enter a capture orbit with a 50 000 km apoapsis. An approximately 6 month aerobraking period lowers the apoapsis to 259 km, with chemical propulsion again used to raise the periapsis into the final circular 259 km altitude science orbit. Sufficient fuel is retained to maintain this orbit within a ~100 m corridor for a 4 year, 6 Cycle science mission (1 Cycle = 1 Venus sidereal day of 243 Earth days) from 8 November 2030 to 5 November 2034.

Addressing the issues raised in the EnVision M4 proposal debrief, the mission is focussed on achieving the science goals rather than global coverage; the proposed spacecraft is simplified, with a fixed 3 m X/Ka-band antenna, a dedicated cold face, and fewer deployable mechanisms; and reduced operational complexity and data volume. All science investigations are carried out in the nadir direction, with the spacecraft rolled by up to 35° during SAR operations.

In conclusion, the EnVision mission takes advantage of Europe's world-leading position in both Venus research and in interferometric radar to propose a mission which will address universally relevant questions about the evolution and habitability of terrestrial planets. In doing so, it will provide a range of global image, topographic, and subsurface data at a resolution rivalling those available from Earth and Mars, inspiring the public imagination and the next generation of European scientists and engineers.

1. *EnVision Science Goals*

New discoveries of Earth-sized planets in orbit around other stars stimulates the need to better understand the planets orbiting our own Sun, particularly those closest and most similar to Earth. In terms of the parameters associated with life, Venus is profoundly alien and Mars the more benign planet, but in geological terms, and in the parameters currently accessible for characterising exoplanets, Venus is the most Earth-like planet in the Solar System. Early Mars may have had limited favourable conditions for life but at one tenth the mass, it was unable to sustain its early benign environment. Being so similar to Earth, Venus may also have had a habitable past, possibly even sustaining a living biosphere. So why has Venus not turned out more like Earth?

This question is tied to our general understanding of the universe and lies at the heart of the Cosmic Vision questions:

- *What are the conditions for planet formation and the emergence of life?*
- *How does the Solar System work?*

Surprisingly little is known about our nearest planetary neighbour, not even the basic sequence and timing of events that formed its dominant surface features. NASA's 1989-1994 Magellan mission provided a global image of the surface at 100 – 200 m resolution, comparable in coverage and resolution to that of Mars after the Viking missions in the 1970s. Magellan revealed an enigma: a relatively young surface, rich in apparent geological activity, but with a crater distribution indistinguishable from random (Figure 1). The initial conclusion was that a global catastrophe half a billion years ago had resurfaced the planet: Venus was solved. After Viking, Mars was similarly thought to be understood, with everything known that needs to be known. Two decades later, Pathfinder reignited public and scientific enthusiasm in Mars and since then newer and higher resolution data from MGS, MRO and Mars Express have revolutionised our understanding of current and past processes alike.

ESA's 2006-2014 Venus Express, the most successful mission to Venus in the last two decades, revealed a far more dynamic and active planet than expected, uncovering tantalising evidence for present day volcanic activity that demands further investigation. Nonetheless, the enigma remains: how can a geologically active surface be reconciled with the global stasis inferred from the apparently random impact crater distribution? The key goals for EnVision are to:

- *Determine the level and nature of current geological activity;*
- *Determine the sequence of geological events that generated its range of surface features;*
- *Assess whether Venus once had oceans or was hospitable for life; and*
- *Understand the organising geodynamic framework that controls the release of internal heat over the history of the planet.*

With its unparalleled European instrument and technology heritage in surface change detection and monitoring, EnVision will revolutionise our understanding of Venus and enable us to understand why our closest neighbour is so different.

Observations from Magellan data imply a variety of age relationships and long-term activity[32,39], with at least some activity in the recent past[56,120,135]. There is a non-random distribution of topography (the highs particularly are semi-linear features) and an association between geological features and elevation, such that the uplands are consistently more deformed than the lowlands. The distribution of impact craters is not strictly random either[24,68,120], with recent observations about the degree of crater alteration[72] permitting a wider range of possible recent geological activity[24,64,66,78,140].

Steep slopes and landslides are very common on Venus, implying active uplift, but existing data provide no constraint on current rates of tectonic activity. The surface of Venus is not organised into large plates like Earth's oceans but it is partitioned into areas of low strain bounded by narrow margins of high strain, analogous to continental basins and microplates. Are these regions actively created and destroyed, like

Earth's oceans, or simply mobilised locally? What is the significance of the global network of elevated rift systems (Figure 3), similar in extent to mid-ocean ridges but very different in appearance? Unique to Venus are coronae, quasi-circular tectonic features, typically 100–500 km across, with a range of associated volcanic features. Are coronae the surface expression of plumes or magmatic intrusions? What role do they play in global tectonic and volcanic change?

**Figure 1       Global Crater Distribution**

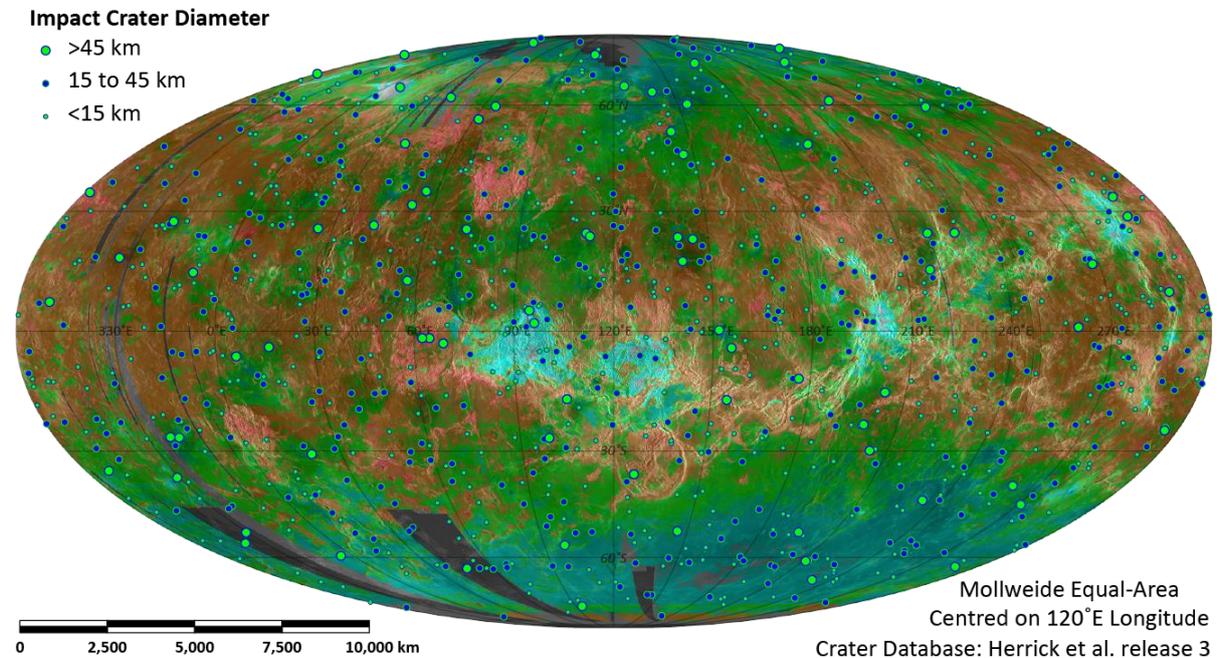

*That the spatial distribution of impact craters is indistinguishable from a random is a puzzle because no other features on Venus occur at random. Underlying colour map shows surface materials: pink – loose sediment; brown – sedimentary or weathered rock; green – volcanic rock; cyan – low permittivity materials.*

Recent and perhaps ongoing volcanic activity has been inferred in both Venus Express[95,131,136] and Magellan[18] data (Figure 2). Maintenance of the clouds requires a constant input of $H_2O$ and $SO_2$[22] which equates to a magma effusion rate of only $0.5$ km³ a⁻¹, assuming a saturated magma source.

However, only one significant volatile-rich pyroclastic flow deposit, Scathach Fluctus[60], has been identified to date, and the morphology of most larger volcanoes is consistent with low volatile eruptions. The actual magmatic rate is likely far higher, ~10 km³ a⁻¹, about one third Earth's[62].

Constraining volcanic activity is critical to understanding when and how Venus was resurfaced, but it is also important to constrain the nature of that activity.

**Figure 2       Thermal anomalies on Venus**

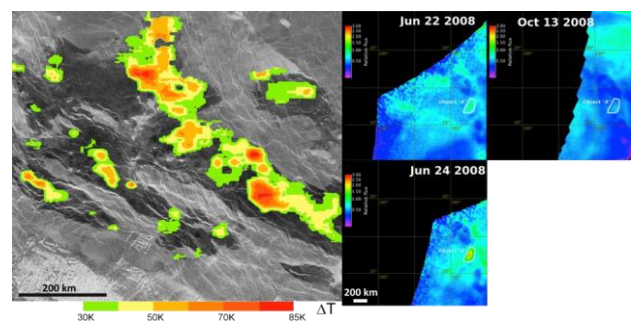

*Left: Magellan thermal anomalies calculated in Bereghinia Planitia, calculated from microwave emissivity data[18]. Right: Maps of relative thermal infrared flux from the surface from Venus Express orbits 793, 795, 906 in Ganiki Chasma[131].*

Are there other large pyroclastic eruptions or is Scathach Fluctus unique? Are canali or other specific magmatic features confined to a past regime or still active today? Is there a correlation between mesospheric $SO_2$ concentration and volcanic activity? Are crater floors effusively infilled and buried from below? Were the plains formed from a few massive outpourings in a short period of time or from many thousands of small flows over their entire history? Or were they formed, or modified, in an entirely different way?

**Figure 3  Volcanic and Tectonic Features**

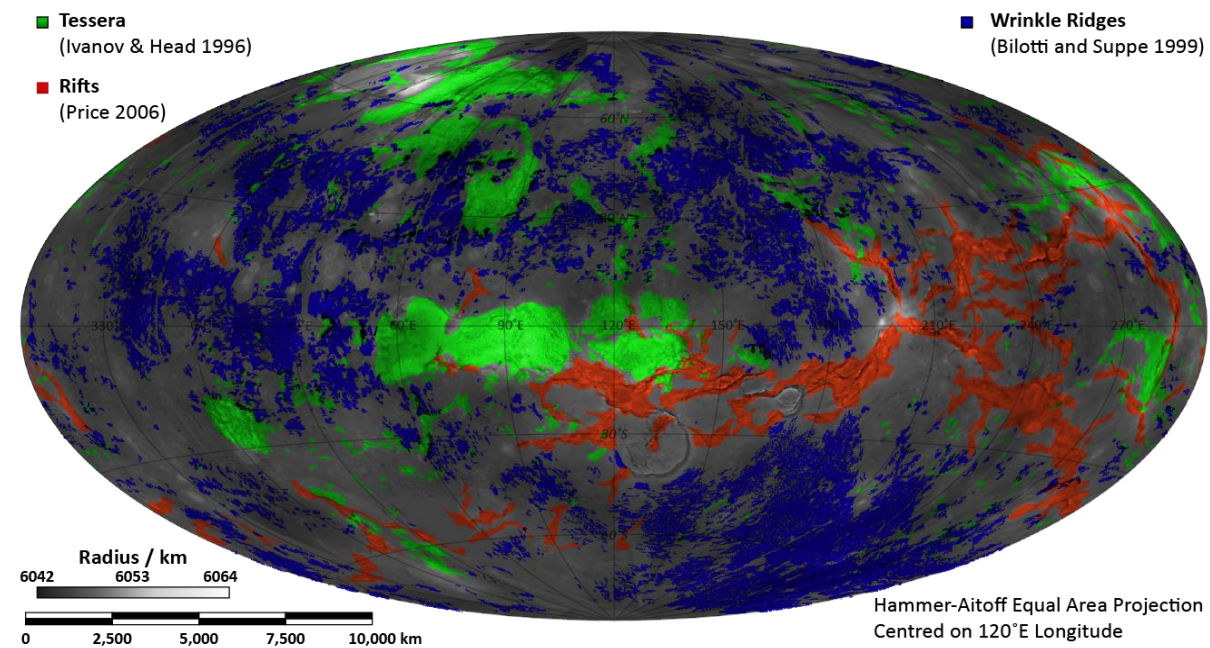

*Rifts follow topographic rises along great circle arcs, similar to Earth's mid-ocean ridges; wrinkle ridges are predominantly in the lowlands. Tesserae are highly deformed terrain across a range of elevation, and are possibly continental crust.*

The slow moving dense lower atmosphere of Venus creates a sedimentary environment similar to the deep oceans on Earth, so that dunes and other aeolian features are rarely large enough to be visible in Magellan images. Understanding modern sedimentary processes is key to distinguishing whether ancient deposits formed under similar conditions or under more benign water oceans.

Surface images captured by Soviet Venera landers (Figure 4) reveal a landscape more consistent with pyroclastic or sedimentary deposits, not the basaltic lava flows widely assumed to cover the plains. The bedrock recorded at the Venera 10, 13 and 14 sites consists of laminated or thinly bedded sheets with varying degrees of coarse sediment or regolith.

Although chemically similar to basalts, the layering is more similar to sedimentary or pyroclastic bedding[47], formed by cycles of air fall or ground flow. Based on load carrying capacities derived from the penetrometer and dynamic loads during lander impact[96,143], the strength of the surface at the Venera 13 site is similar to that of a dense sand or weak rock.

At the Venera 14 and Vega 2 sites the recorded strengths are higher but similar to that of a sedimentary sandstone and less than half that of an average basalt.

A major problem is that almost the entire area imaged by each Venera lander sits within a single Magellan SAR (Synthetic Aperture Radar) pixel, and their landing position is known to only ~150 km, so that it is impossible to correlate features observed in the lander images with those in Magellan images. Do the lander images represent a surface weathering veneer on otherwise intact lava flows, or thick accumulations of aeolian or pyroclastic deposits?

**Figure 4    Venera Landing Sites**

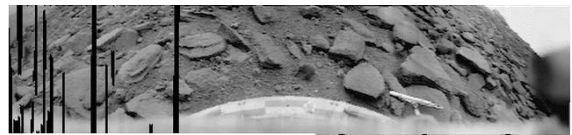
Venera 9

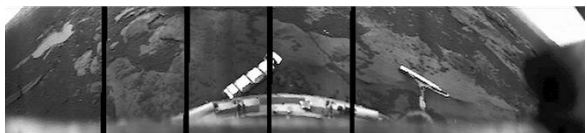
Venera 10

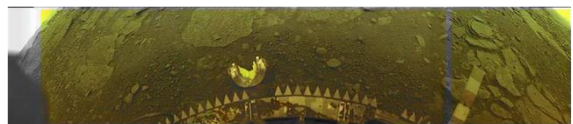
Venera 13

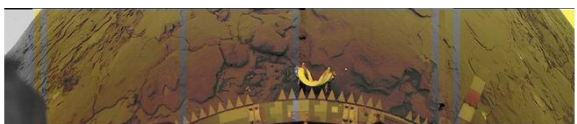
Venera 14

*Venera 9 landed on a talus slope of about 30˚; Veneras 10, 13 and 14 landing on rolling plains with varying amounts of loose sediment and plate-like bedrock[96]. Reprocessed lander image data © Don P. Mitchell, used with permission.*

Nothing is known about the internal properties of Venus: it is less dense than expected if it had Earth's bulk composition but its moment of inertia, the most powerful way to constrain the first order radial structure of a planet, is unknown. Indeed, the shape of the planet appears to be unconnected to its rotational rate, which is too small to explain the observed flattening, but strangely has been shown to vary (Figure 5) by more than 7 minutes in observations throughout the last 40 years[82,106]. The cause of this variability, and whether it is periodic or random, is unknown. The tidal Love number, estimated from Doppler tracking of Magellan and Pioneer Venus Orbiter spacecraft data, indicates that Venus's core may be at least partly liquid[87] but its size is unconstrained. The Venera landers returned a number of K, U and Th measurements that imply bulk ratios, and hence internal radiogenic heating rates, comparable with Earth[109]. Magellan gravity data are consistent with an organised pattern of mantle convection broadly similar to Earth but lack the resolution necessary to understand its connection with geological-scale features, such as individual coronae or mountain belts. Does mantle convection drive geological activity at the present day or is the surface the relic of a past global catastrophe? Does the rotation rate of Venus vary periodically or at random? What does its cause reveal about the interior structure of the planet?

**Figure 5    Spin Rate Variability**

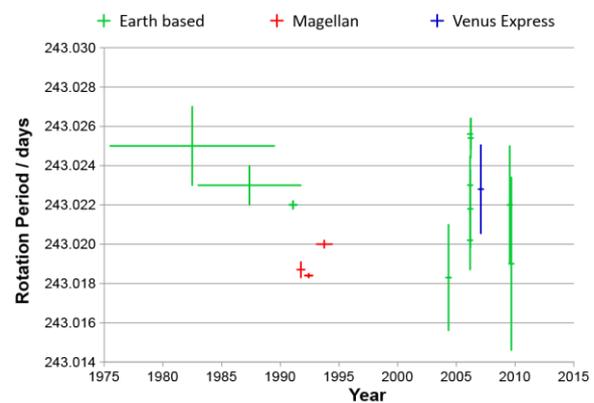

*Venus apparently rotated more quickly during the period of the Magellan mission (small red error bars, 1990-1992) than it did in the first Earth-based observations (green) or in later measurements from Earth and by Venus Express (blue). Vertical bars indicate measurement uncertainty, horizontal bars the period over which the measurement was made.*

EnVision is designed to answer all these diverse questions using a simple but complementary suite of instruments: VenSAR, a 3·2 GHz phased array synthetic aperture radar; SRS, a 16 MHz subsurface radar sounder; and VEM, a Venus Emission Mapper comprising VEM-M (infrared emissivity mapper), VEM-H (high spectral resolution infrared spectrometer), and VEM-U (ultraviolet spectrometer). The spacecraft's telemetry systems comprise an additional instrument for the purposes of gravity-field determination and occultation studies.

## 2. Science Requirements

The three key science goals are related to processes operating at different spatial and temporal scales and hence EnVision requires a complementary suite of observations to address them. In detail, the processes of interest are global-scale processes operating over the lifetime of the planet that sustain regional tectonic and volcanic processes operating over the observable geological history. The nature of these processes and whether they occur episodically or in a steady-state is keenly debated; to distinguish between them requires a better understanding of how global processes drive individual features, such as volcanoes, rifts, and mountains, and the proportion of these features that are active at the present day. Achieving this requires not a global overview like that provided by Magellan but rather, a detailed and comprehensive assessment of a representative subset of these features.

Geological processes operate at all scales, as recognised in conventional mapping[99] (Table 1). This hierarchy differentiates processes that operate at, and affect, features at the different scales indicated, and requires a resolution at least 2-3 times finer to discriminate these features. The Zonal-scale 100–200 m resolution of Magellan imagery enables mapping of the global distribution of volcanoes, for example, but not their age relationships, which would require Reconnaissance-scale imaging to reveal the cross-cutting relationships between different flows.

*Table 1        EnVision Mapping Hierarchy*

|  | *Global* | *Zonal* | *Reconnaissance* | *Exploration* | *Locality* |
|---|---|---|---|---|---|
| Coverage | >95% | >95% | >20% | >2% | >0·2% |
| Unit Area | Global | 2500 × 2500 km | 1500 × 1500 km | 100 × 100 km | 5 × 5 km |
| Resolution | 50 km | 150 m | 30 m | 6 m | 1 m |
| Feature Size | 150 km | 500 m | 100 m | 20 m | <4 m |
| *Geomorphological Features* | | | | | |
| Structures | Terra 'continents', Planitia | | Chasmata, Dorsa | Folds, graben | Fault scarps |
| Volcanoes | | Volcanic rises (Regio) | Volcanic edifices | Lava Flows | Flow textures |
| Sediments | | 'Featureless' plains | Parabolas, halos | Landslides | Dunes |

*Geological processes operate across a range of scales; while global metre-scale data would perhaps be ideal, the data volume would be prohibitive to return and analyse. Instead, a nested set of observations sampling decreasing areas at increasing resolution, are sufficient to characterise the processes involved, e.g. textures observable at the Locality (metre) scale help to understand flows at Exploration scale, which help to understand edifices at the Reconnaissance and rises at the Zonal and Global scales.*

In Earth observations, ERS-1, ERS-2 and ENVISAT all provided 30 m resolution data (the latter also 150 m and 1000 m data). COSMO-SkyMed offers 100 m, 30 m and 5 m stripmaps and 1 m spotlight images and TerraSAR-X/Tandem-X 3 m stripmaps and 1 m spotlight images. Sentinel-1 data are available from 5 m to 40 m resolution. Not only have these resolutions proved effective on Earth, adopting the same resolutions on Venus means that there will be a wealth of comparable data from Earth. Indeed, NovaSAR will acquire 30 m and 6 m stripmap imagery at the same frequency, providing data directly comparable with EnVision.

Conceptually, therefore, EnVision is designed to deliver nested data[153], from measurements of the gravity field, spin rate and axial wobble at the global-scale, to metre-scale observations of current rates of activity and stratigraphic relationships, and thence on to a selection of locality-scale snapshots to show how global change is effected, from the smallest scales upwards.

Before discussing how this is accomplished, this section expands on the detailed science requirements of the mission, on the basis of which EnVision's instruments were selected.

### 2.1. *Global Scale Dynamics*

Venus rotates so slowly (its sidereal day is 243 days long and is commonly called a Cycle) that its flattening is unrelated to rotation and its rate of precession too slow to estimate its moment of inertia. However, Doppler tracking of Magellan and Pioneer Venus Orbiter spacecraft was sufficient to estimate its tidal Love Number ($k_2 = 0.295 \pm 0.066$), which indicates an at least partly liquid core[87]. Cosmochemical models[11,105] suggest core mass fractions between 23.6 and 32.0% — implying a mantle mass proportionately similar to or greater than Earth's. The Venera landers returned a number of K, U and Th measurements that imply bulk ratios, and hence internal radiogenic heating rates, comparable with Earth[109]. Estimation of the maximum amplitude of the radial displacement induced by the solar tides with an accuracy better than 2 cm over half a Venus solar day (58.4 days), together with the time lag of the bulge with an accuracy on the order of 1 hour, would discriminate between different compositional models, as well as providing average mantle temperature and viscosity. Reducing the uncertainty in $k_2$ to ± 0.01 would also distinguish between these compositional models and constrain the thermal state of Venus[42].

Many values of the mean rotation of Venus have been estimated since 1975 from which the length of day has been shown to vary by more than 9 minutes. Analysis of several mechanisms that can induce oscillations in the rotation rate, including triaxial coupling, tidal deformation, atmospheric coupling and core/mantle coupling, found that the instantaneous rotation rate could vary by as much as 3 minutes[35], principally from triaxial coupling. This value corresponds to a variation in the longitude of a reference point by 12 m at the surface within a Venus solar day (116.8 days)[82]. Repeated determination of the rotation rate and orientation of the spin axis[37] at least several times each Cycle (1 Cycle = 1 Venus sidereal day = 243.02 days) are required to determine any periodic effect in the rotation rate. In addition to precise orbit determination, EnVision will achieve these measurements by tracking the locations of at least four Venera landers twice per cycle using high resolution SAR images and obtaining the orientation of the polar spin axis from repeated interferometric SAR (InSAR) images of both polar regions. Interferometric measurement of polar axis was demonstrated by Magellan (Figure 10), with an uncertainty below 15 arcsec (Goldstein, *pers. comm.*). From these measurements the amplitude of the nutation can be determined, which with the gravity coefficient may be used to infer the moment of inertia. In addition, orbit reconstruction might also place constraints on the length of day measurements, as it has for Mars[86,88].

Linking the global interior dynamics to lithospheric processes and structure is the relationship between topography and gravity. Depending on the mode of topography compensation, the gravity and the topography signals exhibit higher or lower admittance and correlation. Magellan tracking data used to construct the current gravity field model were obtained during Cycle 4, from a highly elliptical orbit (170 × 14 500 km), and during Cycles 5 and 6 when the orbit had been circularised to 197 × 541 km. Since the local resolution depends primarily on altitude and data can only be obtained when the spacecraft is transmitting with line-of-sight to Earth, the degree strength varies from degree 100 to as low as degree 40, equivalent to a spatial resolution varying from 190 km to 475 km. Low degrees of the gravity field correspond mainly to large-scale internal dynamics while intermediate degrees (up to at least 80) are sensitive to lithospheric structure and compensation processes[5,114]. The existing gravity field resolution is consequently too low to study lithospheric structure and compensation processes of many features of interest, including the potentially active hot spot identified in Venus Express VIRTIS data, and smaller-scale structures such as coronae. These 300~500 diameter features are probably caused by small-scale mantle upwelling and/or intrusions within the lithosphere. To understand such structures a spatial resolution of <200 km or better is required in both gravity and topography, equivalent to a spherical harmonic gravity field known to at least degree and order 120.

Gravity data reveal the mantle processes driving regional tectonics on Venus but how are those processes organised and translated into features at the surface, and how active are those features? EnVision will use its full complement of instruments in concert to find out.

### 2.2. *Regional Processes*

The lack of plate tectonic features such as spreading ridges and subduction zones; the close correlation between geoid and topography at both long and short wavelengths[133], unlike Earth; and the near random distribution of the ~940 impact craters on Venus, imply a stagnant lid regime[8,122,123,138] and a globally uniform surface age[101,117,142] of ~750 million years. A proposed global stratigraphic sequence[12,74] suggests rapid global resurfacing, probably episodic[52,110,111,148,149], followed by a long period of quiescence. However, observations from Magellan data reveal an array of organised geological complexity[5,58,77] implying a variety of age relationships and long-term activity[32,39], at least some of which was in the recent past[57,120,137]. There is a non-random distribution of topography[50,121], deformation[81,141] and volcanism[70]; the distribution of impact craters is not strictly random either[24,68,120], with recent observations about the degree of crater alteration[72] permitting a wider range of possible recent geological activity[64,66,78,140]. While tesserae on the highland plateaus and elsewhere may be the equivalent of continental crust on Earth, they cover only a quarter as much area and must occur across a wider range of elevations, since Venus lacks Earth's bimodal topography[20,61,67,125,126,132].

These features partition the Venus surface into regions, referred to as terranes of low strain surrounded by narrow belts of high strain (Figure 6; note that this term has a narrower usage in terrestrial geology), but unlike Earth's plates they are typically only 500~1500 km across, the same order of magnitude as the ~800 km average crater spacing, and so likely important in understanding both the crater distribution and global resurfacing.

**Figure 6       Example Terrane**

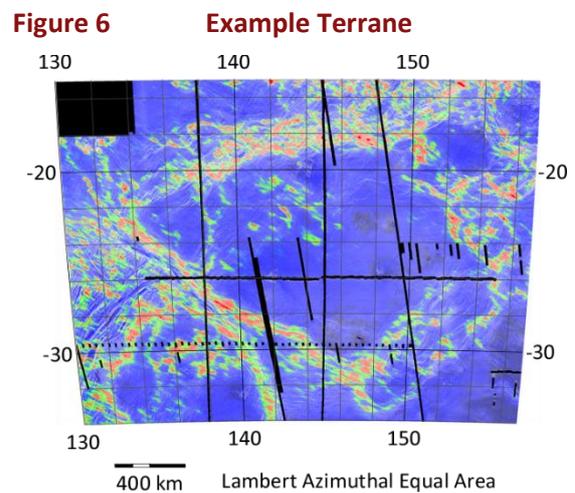

*Magellan SAR image with false colour regional slope, showing an average-sized tectonic terrane comprising an undeformed interior (blue) surrounded by relatively diffuse deformation belts (green to red). Notice that these outline neighbouring terranes.*

Terranes on Venus have a wide variety of morphologies ranging from, for example, the 600 km diameter Atete Corona to the 1500 km tessera plateau of Alpha Regio. Understanding their nature – how they are deformed and reworked – is therefore crucial to solving the paradox between the geological complexity of Venus and its crater distribution. The key questions to be answered are:

- What is the connection between these terranes and underlying mantle convection?
- How rapidly are the high strain margins being deformed and by what processes?
- What processes modify the low strain interiors and over what timescales?
- Are there distinct compositional differences between terranes?
- What is the relationship between terranes and volcanic processes?

Addressing these questions requires the improved gravity field discussed earlier and a range of complementary observations to

### 2.2.1. Radar Mapping

Differential InSAR (DInSAR)[98] is the only tool capable of measuring geological-scale strains from orbit and is particularly effective across high strain rate terrane margins, in which LoS displacements may be 10 mm a$^{-1}$ or more. Combining LoS displacements derived from D-InSAR sets in ascending and descending (opposite look) orbits allows the vertical and at least one of the horizontal components of displacement to be isolated[73,155]. Two complementary methods[13,44] are commonly used to detect displacements as small as 1 mm a$^{-1}$, even in the absence of an earthquake[89]. Combining these techniques with opposite look sets to isolate components of movement means that even the low strain deformation of terrane interiors is detectable with DInSAR[59,97].

Many fracture sets visible in Magellan images appear to have formed in response to subsurface dykes[100,112], which often occur in swarms that radiate in patterns related to the global stress state of the lithosphere[63]. Coronae[10] – unique to Venus – also appear to be the surface expression of subsurface intrusions or magmatic plumes[139] and recent research[103] suggests that intrusions may be more important on Venus than Earth because its weak lower crust[7] inhibits extrusion[91]. DInSAR is highly effective at detecting magmatic inflation under terrestrial volcanoes[17,51], even where no volcanic feature is evident[151], making it the ideal tool to study magmatic processes associated with terrane margins and interiors. DInSAR may therefore reveal whether different rift morphologies and corona associations are related to an increased rate of subcrustal stretching and intrusive magmatism[58] or to different rift ages[108,135], and hence illuminating the details of the connections between surface features and underlying mantle processes.

Coherence, a by-product of DInSAR, is also useful for change detection[146]: a reduction or loss of coherence implies change at the scale of the radar wavelength or above. Atmospheric effects, particularly changes in the cloud layers, are the primary factor in loss of coherence but are long wavelength features (at least 50 km) that can be corrected for[40]. Surface changes causing coherence loss are usually smaller in scale and geographically distinct, often in the form of channels and lobate downslope mass movements[129]. Canali are river-like channels thought to be formed by carbonatite or sulphur-rich volcanic flows[83,85,152] or sedimentary density currents[150]; coherence data will distinguish between these possibilities from their pattern of coherence loss. Mass wasting[49] and landslides are common on Venus[92] and may contribute to a small but global supply of sediment, revealed in Magellan Doppler Centroid data[19] and Venera lander images[48]. Thus the pattern of coherence loss can reveal other mechanisms of surface change in addition to those from tectonic or volcanic processes.

Coherence and DInSAR change detection only reveal current rates and styles of activity and not whether these are in a long-term steady state, in gradual decline, or a lull between episodic global resurfacing events. Worse, even steady-state processes may appear infrequently and episodically on an annual to decadal timescale. To fully understand the behaviour of the Venus lithosphere over time requires geological mapping to ascertain stratigraphic relationships and hence geological history[65]. This requires a knowledge of the geological materials at a resolution sufficient to distinguish between stratigraphic units.

Radar is sensitive primarily to the morphology (roughness and slope) and relative permittivity of the surface materials. Polarimetric data provide important information about the nature of the surface and near subsurface that cannot be obtained solely with backscatter power images, such as those obtained by Magellan. In particular, polarisation ratios can help identify the thickness and grainsize of loose surface sediment[53]. Since terrestrial studies show that almost all natural targets have reciprocal cross-polarisation (i.e. HV backscatter is identical with VH)[145], only HH, VV, and VH (or HV) polarisations are required to characterise the backscatter properties. Arecibo data have demonstrated the utility of this at Venus for distinguishing volcanic deposits[28], impact ejecta[26] and a thin, patchy but widespread regolith[29] consistent with Venera lander images, but these data are at a resolution of 12 to 16 km, too low to discern detailed stratigraphic and geological relationships.

The relative permittivity (also called the dielectric constant) of near-surface materials can be inferred from their microwave emissivity, itself a derivative of the radar brightness temperature measured by using the SAR antenna as a radiometer. In this mode the resolution is dependent on the real antenna size and hence is very low, e.g. Magellan emissivity data have a resolution of ~50 km, but a larger antenna can reduce this to a few kilometres. For most natural materials the relative permittivity depends upon density and can be used to distinguish between areas of loose sediment, weathered rock and exposed fresh rock[27]. Certain materials, e.g. metals, have very high relative permittivity which lowers the emissivity, making these materials very bright in radar imagery. On Venus, slightly elevated relative permittivity occur in certain volcanic materials, probably Ti-rich basalts[60]; parabolic ejecta halos may have low or moderately elevated relative permittivity[23]; and very high relative permittivities occur at most, but not all[25], high elevations[9,116]. The cause of these very high relative permittivities is unknown and require polarimetric data, and observations at different wavelengths, to understand their origin.

#### 2.2.2. Subsurface Mapping

Putting these data together into a geologically-meaningful context requires knowledge of the third dimension. Topography can be acquired from orbit in three ways: from radar sounding (altimetry); radargrammetry (from stereo pairs and shading); and from InSAR. The latter is normally a necessary step in the production of DInSAR data but while relative shifts of a few mm are readily detectable, absolute elevations depend on a number of factors, such as orbit knowledge, that limit the vertical resolution to ~15 m. For typical stereo separations of ~20° and automated matching, radargrammetry provides a similar vertical resolution but at a lower spatial resolution. These two techniques can be considered complementary, since steep slopes prevent InSAR by causing loss of coherence, but automated stereo matching is most effective in rough terrain; and both can be derived as a by-product of data acquired for other purposes.

Radar sounding can provide continuous profiles at a resolution limited only by orbit knowledge and ground track spacing, which in the case of EnVision is 10 km at the equator. Operating at a frequency below ~30 MHz has the further advantage that the signal penetrates the ground[80,130], providing information on subsurface structures that are crucial to understanding the history of Venus. The two end-member hypotheses, episodic/directional[12] and equilibrium/non-directional[64], predict that the plains comprise lava flows that are predominantly extensive and thick (~100s m), or local and thin (~10s m), respectively. A sounder in the 9-30 MHz range is able to penetrate to a crustal depth of 750–340 m respectively and image subsurface features at a vertical resolution of 5-16 m, more than adequate to distinguish between these end-member hypotheses, as well as providing topographic profiles.

#### 2.2.3. Infrared Mapping

Radar alone cannot distinguish between different rock compositions. On Mars and other planets, infrared reflectance spectra offer the most effective way to determine composition from orbit. There are spectrally useful night side infrared windows at 0·8 to 1·18 μm in the otherwise opaque Venus atmosphere[2,3,36] but scattering in the global cloud layer limits their spatial resolution to ~50 km. Galileo and Venus Express data have shown that highland tessera terranes have lower, and many volcanic provinces higher, infrared emissivities than the global average[67,107], implying compositional differences. More precise measurements are required across all the available spectral windows in order to identify the mineralogical differences between these terranes. Repeated imagery of thermal emission from the surface at 0·8 to 1·18 μm may reveal time variable signatures characteristic of volcanic activity, following tantalising hints from Venus

Express[131], is useful for establishing current levels of volcanic activity on Venus.

In summary, understanding regional processes requires a range of complementary observations across a representative sample of different terrane types, with each survey built up from individual swaths into regions 1500 × 1500 km across. Where not constrained by natural limitations (e.g. cloud scattering) these observations should be at a resolution able to distinguish regionally important rock units and their relationships sufficient to understand the geological history of each terrane, e.g. flow units rather than individual flows. However, geological processes operate at all scales, and geological change is often the result of incremental small-scale processes. DInSAR and coherence data will show where and when these changes occur but to understand them requires much higher resolution observations.

### 2.3. *Small-Scale Change*

Terranes host a variety of discrete features – individual lava flows, faults, landslides, etc. – that link global processes directly with small-scale change. The identification of new lava flows, particularly smaller flows, places a tight constraint on the nature and rate of volcanic activity[90]; similarly the number and size of new landslides constrains the frequency and scale of seismic events[92]. Understanding processes at the metre scale is therefore critical to understanding global activity and change. Geomorphology at this scale is essential for understanding processes of mass wasting – landslides and talus slopes – that are indicative of geologically active slopes and tectonic activity. Boulder tracks, in particular, can be used to estimate the source and magnitude of earthquakes[124], complementing D-InSAR data.

Highland plateau tesserae may be continental terranes with long and complex histories[32,57,67,125,127]. Magellan images show hints of stratigraphic layering; unravelling that history requires detailed mapping at the highest resolution. Details in layering on scarp slopes may reveal the chronology of resurfacing and the nature of past environments, and be usefully correlated with subsurface features, helping to reveal whether Venus supported oceans[79] and perhaps, therefore, life, early in its history, or whether indeed it was hotter[128] in the past.

Large dunes are very rare on Venus, with only two dune fields identified in Magellan imagery, both related to impact cratering. However, Venera 13 imaged active wind transport and ripple-like features[48]; putting these observations into a wider context requires imaging of the lander sites at the metre-scale. Correlation with a variety of other sites across Venus is needed to understand the nature and importance of aeolian processes, the role they play in the exchange of volatiles between the atmosphere and interior, and the stability of geochemical cycles on Venus.

There is therefore a need for a detailed survey at the Exploration scale (~6 m resolution), about a factor of five better than the 30 m resolution Reconnaissance surveys, across a representative selection of features within each terrane. The areas selected should be ~100 km across and focussed particularly on those areas identified in the Reconnaissance survey as either active or ancient.

### 2.4. *Surface-Atmosphere Exchange*

Nowhere is the link between the local scale and global processes more obvious than in the effects of a single volcanic eruption on global atmospheric chemistry and climate. These effects can be significant on Earth but no eruption has been directly observed on Venus. Dramatic shifts in mesospheric $SO_2$ levels have been detected[43,94] that may be reflect eruption events but cannot be distinguished from other atmospheric processes[93] without measurements in the lower atmosphere linked to specific localities. Maintenance of the clouds also requires an input of $H_2O$[22], most likely also from volcanic sources.

To calculate whether a volcanic eruption could be detected using infrared sounding from a satellite, the likely compositional changes that would result can be estimated. The nominal column mass of volcanic gases in the Venus atmosphere, integrated from surface to space, is ~200 kg m$^{-2}$ for $SO_2$, ~10 kg m$^{-2}$ for $H_2O$ and

~0·1 kg m⁻² for HDO[38]. If the composition of Venus volcanic gases is the same as on Earth and provided that plume dispersion does not exceed $10^4$ km², the limiting spatial resolution induced by cloud scattering, then a large, Pinatubo-size eruption would alter the composition in the following way:

- Increase $H_2O$ by several tens of percent;
- Decrease D/H ratio by several tens of percent; and
- Increase $SO_2$ by about 1%.

The latter effect is probably underestimated with respect to the others, since the Venusian interior is thought to be much drier than Earth's, so that the outgassed $SO_2/H_2O$ ratio may be much higher on Venus. Observations of changes in lower atmospheric $SO_2$ and $H_2O$ vapour levels, cloud level $H_2SO_4$ droplet concentration, and mesospheric $SO_2$, are therefore required to link specific volcanic events with past and ongoing observations of the variable and dynamic mesosphere, to understand both the importance of volatiles in volcanic activity on Venus and their effect on cloud maintenance and dynamics.

The spectral window at 1·18 μm probes the first scale height of Venus' atmosphere, which directly interacts with the surface. The high atmospheric pressure at this level widens spectral lines so that a moderate spectral resolution of a few cm⁻¹ between 1·08 and 1·20 μm is sufficient to resolve individual $H_2O$ and HDO lines to measure both water vapour abundance and isotopic ratio. Sensitivity tests using the appropriate radiative transfer model[16] indicate that a SNR of about ~100 is required.

The 2·4 μm spectral window probes higher in the atmosphere, at 30 to 40 km, but gives access to additional minor species, including $SO_2$ in the 2·450 to 2·465 μm spectral interval and OCS (carbonyl sulphide) in the 2·440 to 2·465 μm spectral interval. Individual lines need to be resolved in order to distinguish between all species, requiring a very high spectral resolution of more than 40 000 between 2·44 and 2·47 μm. Sensitivity tests using the radiative transfer model for this altitude range[119] show that a SNR better than a few hundred is required for an accuracy of 1% on $SO_2$ retrievals. Hence the spectral requirements are:

- 1·08–1·2 μm at R = 2000 ($H_2O$, HDO at 0–15 km)
- 2·44–2·47 μm at R = 40 000 ($H_2O$, HDO, OCS, $SO_2$ at 30–40 km)

Linking the lower atmosphere volcanogenic volatile variability with the known variations[43,94] of cloud top column in SO, $SO_2$ and blue-UV absorber leads to the following specifications in the UV range:

- 200-230 nm at 0·3 nm spectral resolution to distinguish SO spectral lines from $SO_2$[76]
- 170-400 nm at 1·5 nm spectral resolution to fully encompass the UV absorber signature and provide (SO+$SO_2$) measurements.
- 0·1° angular resolution, equivalent to about 300 m at cloud top level.

For change detection the temporal coverage should be as frequent as possible, ideally once per Cycle, at a spatial resolution of a few hundred kilometres.

A key goal for the detailed survey mode is to determine the exact location of the ten Soviet automatic landers (Veneras 7–14 and Vegas 1 and 2), and possibly the remains of the US Pioneer Venus Large Probe, within their landing ellipses, in order to use them as geodetic control points. These probes will appear as very bright points ~25 dB brighter than the surrounding plains in the detailed surveys.

Four of these automatic landers (Veneras 9, 10, 13 and 14) successfully returned images from the surface of Venus[48,49,54] but each panoramic scene returned covers no more than a few pixels of Magellan imagery[1] and so lack any wider context (

Although chemically similar to basalts, the layering is more similar to sedimentary or pyroclastic bedding[47], formed by cycles of air fall or ground flow. Based on load carrying capacities derived from the penetrometer and dynamic loads during lander impact[96,143], the strength of the surface at the Venera 13 site is similar to that of a dense sand or weak rock.

At the Venera 14 and Vega 2 sites the recorded strengths are higher but similar to that of a sedimentary sandstone and less than half that of an average basalt.

A major problem is that almost the entire area imaged by each Venera lander sits within a single Magellan SAR (Synthetic Aperture Radar) pixel, and their landing position is known to only ~150 km, so that it is impossible to correlate features observed in the lander images with those in Magellan images. Do the lander images represent a surface weathering veneer on otherwise intact lava flows, or thick accumulations of aeolian or pyroclastic deposits?

Figure 4). The panoramas reveal a variety of surface materials, including loose regolith and layered rocks, indicative of weathering and sedimentary processes. Understanding the nature and extent of these surface materials requires stereo polarimetric data at the Locality scale, ~1 m resolution over a few kilometres, to determine local topography and boulders, and distinguish bare rock, loose sediment, boulders, and other materials.

Linking radar data to the lander observations in this way is extremely useful in understanding materials at localities elsewhere. A representative selection of Locality scale snapshots, each a few kilometres across, from within areas imaged at Exploration scale (6 m resolution), are needed to fully interrogate active or ancient sites and features that poorly understood, such as farra (pancake domes) and canali.

The nested survey approach, using a range of complementary observations, has proved highly successful on the Moon and Mars and will doubtless transform our understanding of Venus. How this is achieved within one mission is the subject of the next section. In conclusion, the geological investigation of Venus requires investigations of target regions at different spatial resolutions and with different types of observations, from imagery and polarimetry to topography, at a range of scales from 1 m to 30 m, to complement Magellan radar maps, and both infrared and microwave radiometry.

# 3. Proposed Scientific Instruments

## 3.1. VenSAR

The primary instrument carried by EnVision is VenSAR, a 5·47 × 0·60 m phased array synthetic aperture radar (SAR) antenna, operating at 3·2 GHz, in the S-band, similar in frequency to Magellan's SAR. The opacity of the Venus atmosphere at radar wavelengths approximates a frequency squared dependence and in a practical sense is opaque above 10 GHz. More critically for D-InSAR change detection is the variability in refractive index of the atmosphere, which causes phase shifts in the transmitted and received radar signal. The ionosphere introduces phase shifts that are larger at longer wavelengths. However, the lack of magnetic field means that the total electron count at Venus is only a few percent of Earth's, varying from ~1·6 TECU on the dayside to 0·8 TECU at night. The phase shifts are therefore small and readily corrected for.

Of greater significance are the phase shifts caused by variability in the concentration of sulphuric acid droplets in the cloud layer, with both altitude and latitude, over timescales of several hours to a few days. Although the refractive index is independent of wavelength, the phase shifts are not. Based on Magellan radio occultation data[75], these shifts can cause significant phase ambiguities at frequencies above ~5 GHz (C-band), driving the choice towards lower frequencies. Cycle to Cycle (243 day separation) interferometry has been demonstrated with Magellan data[71] and topographic fringes identified in Arecibo data at the same frequency[30]. However, lower frequencies are less sensitive to surface displacements and hence less able to distinguish between the different models of geological activity. A frequency in the S-band of 3·2 GHz (9·4 cm wavelength) provides a good compromise between these competing factors.

VenSAR is based on the NovaSAR-S antenna technology developed by Airbus Defence and Space, which itself is built on the heritage of Sentinel-1 and ENVISAT but incorporating significant technical advances, particularly the use of mature GaN technology[104] high power solid state power amplifiers. This technology means that VenSAR requires fewer than 1/6th the number of phase centres that would be needed with Sentinel-1 era GaAs technology and is the primary reason for its reduced mass and cost. The microstrip patch phased array provides a self-contained front-end by mounting the RF electronics on the reverse side of the antenna panel. The VenSAR antenna consists of 24 phase centres, in a 6 × 4 arrangement of centre-fed sub-arrays each of which contains 24 patches. Each sub-array is individually controllable in phase, polarisation for transmit (Tx) and receive (Rx) functions, and Rx gain, with a beam control unit to apply transmit and receive phase adjustments. These provide the antenna with considerable flexibility in the selection of resolution and swath width, within the available 182 MHz bandwidth, and incidence angles from 20˚ to more than 45˚.

In normal stripmap SAR operations, the ultimate spatial resolution along track (azimuth) is nominally half the antenna length, ~3 m, while across track (range) it is controlled by the available RF bandwidth, which at 182 MHz is 1~2 m, depending on incidence angle. Radiometric resolution increases with the number of looks but at the expense of spatial resolution. A good compromise is ~9 looks[154]; Magellan images were typically only 5–6 looks in the lower latitudes. At high resolution, therefore, optimal images have 6 looks (2 in azimuth and 3 in range) and a spatial resolution of 6 m, suitable for Exploration scale mapping. SAR resolution is given in the traditional sense of a point-spread function and not metres per pixel, as is usually the case for planetary cameras; SAR pixels are typically about two-thirds the resolution, i.e. 4 m pixels for a 6 m resolution image.

Operating at the full bandwidth has very high power demands (~2 kW at a 20% duty ratio) and data rates (~900 Mbits s⁻¹) that are not required for standard Reconnaissance mapping. By obtaining 9 looks in azimuth, the range resolution can be reduced to ~27 m, ideal for Reconnaissance mapping and requiring only ~15·5 MHz bandwidth, reducing the data rate to

~65 Mbits s⁻¹ and the duty ratio to 4% (~600 W). These data have four times the spatial and three times the radiometric resolution of Magellan; at full resolution the spatial improvement is a factor of 20, for the same radiometric resolution as Magellan. In spotlight mode the spatial resolution is 120 times Magellan, at about half the radiometric resolution.

### 3.1.1. *VenSAR Design*

The VenSAR design is derived from the NovaSAR-S instrument that is currently being built for the UK Space Agency. NovaSAR-S comprises two major sub-systems, an active phased array antenna sub-system (the front-end) and a central electronics sub-system (the back-end). In NovaSAR-S the active antenna is configured from an array of 18 identical phase centres each comprising a 2 × 2 array of dual polar, 6-element sub-arrays which are excited by three distinct equipment units, coupled together with associated wiring harnesses: a transmit (Tx) unit capable of delivering 115 W peak RF power, a single channel low noise amplifier (LNA), front end receive (Rx) unit, a beam control unit, and a power conditioning unit[34]. RF signal distribution networks deliver signals to and from the central electronics sub-system which forms the radar backend. All of these equipment units and sub-systems have been designed, tested, qualified, and manufactured, and are in place on the NovaSAR platform for a launch currently scheduled for late 2016. Hence by 2017 their TRL will be at level 8/9.

The VenSAR design takes the fundamental active phase centre technology (2 × 2 array of sub-arrays coupled with the associated electronics) and configures the active antenna as an array comprising six columns of four rows of NovaSAR-S phase centres. Thus, the technology of the sub-arrays themselves will be at TRL 9. A development programme to bring the physical structure of the VenSAR antenna to TRL 7 is envisaged during the Phase A and B1 so that the antenna stack will be in position to demonstrate TRL 8 at the end of Phase C/D.

Calibration paths have been included to enable characterisation of the phase centre distortions for replica generation, antenna beam pattern maintenance, and system diagnostics. The calibration scheme is based on the scheme developed for ASAR on ENVISAT, with a P1 path that includes the transmit electronics but bypasses the receive electronics, a P2 path that includes the receive electronics but bypasses the transmit electronics, and a P3 path that bypasses both the transmit and the receive electronics. The P1 and P2 paths each have an H and a V variant.

**Figure 7    NovaSAR at Airbus, Portsmouth**

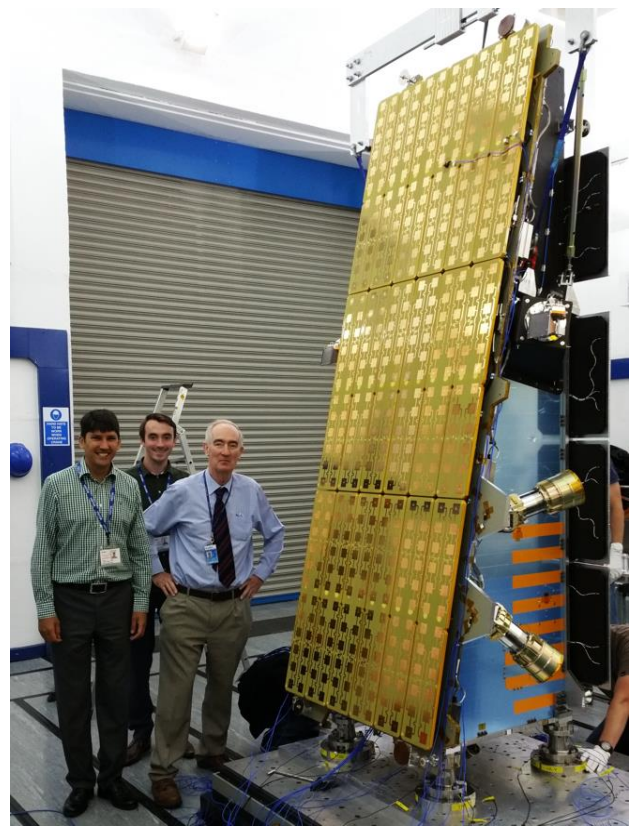

*The NovaSAR-S phased array consists of 3 × 6 phase centres (gold panels); VenSAR will comprise an array of 6 × 4 phase centres.*

The New Instrument Architecture (NIA) generic space radar central electronics exploits the power and flexibility of the Xilinx Virtex 5 (XQR5V) Field Programmable Gate Array (FPGA). The XQR5V is the first high performance RAM based FPGA to integrate effective single event effect mitigation into its core architecture. This has created the opportunity to develop a truly generic backend solution that can easily be applied to a very wide range of space radar missions with minimal non-recurring cost in a compact, lightweight and low power module[33] and is ideally suited to the VenSAR instrument.

Within the front-end, the sub-array radiator assemblies together with the associated passive transmit and receive feed networks consisting of passive Wilkinson style splitters interconnected with coaxial cables are attached by isostatic mounting blades to the outer face of a 25 mm aluminium honeycomb panel.

Table 2      VenSAR mass and power

| Component | Mass | Power |
|---|---|---|
| 2 × NIA Central Electronics | 24·0 kg | 51 W |
| 24 × Front-end Electronics | 104·7 kg | 290 W |
| 24 × Radiator Units | 25·1 kg | 1874 W[†] |
| Antenna Structure | 22·1 kg | |
| Total | 175·9 | |

[†]at 20% duty ratio.

An RF transparent sunshield over the front surface of the antenna serves to reduce the temperature excursions seen by the panel. The RF units are mounted on the inner (satellite) side of the honeycomb panel and are covered with multi-layer insulation to thermally isolate them from the rest of the spacecraft.

The high albedo of Venus means its infrared temperature is only 228·5 K[69], ~30 K cooler than Earth, sufficiently cold for the antenna to radiate the thermal pulse generated by radar operations during the remainder of the orbit, even with reflected solar infrared. Parametric studies (Figure 8) show that the limiting constraint on radar operations is data volume, not the ~310 K operating thermal limit. VenSAR could operate for 5 minutes at 20% Tx and more than 16 minutes at 4%. The thermal pulse at 20% Tx is larger than during aerobraking, when the antenna is aligned parallel to the airflow, leading to confidence that antenna will be able to withstand the thermal loads during aerobraking.

The estimated total mass of the VenSAR front-end is 152 kg, including margins; the component mass and peak power consumption budgets are listed in Table 2. Note that although the radiator units are not planned to operate at transmit duty ratios of more than 20%, they can tolerate a transmit duty ratio of up to 30% for short periods, thus enabling a 1·8 dB improvement in sensitivity if required, for instance when operating in the very high resolution Sliding Spotlight mode.

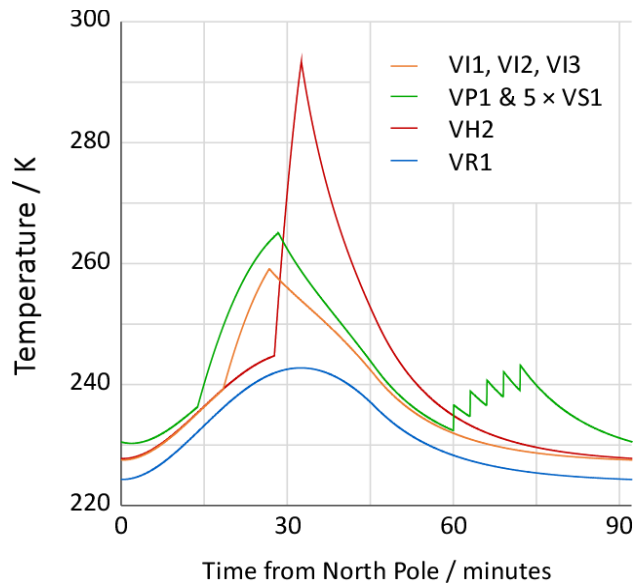

Figure 8      VenSAR Parametric Thermal Analysis (*worst case*)

*Parametric model assumes visible albedo reflection coating and thermal shielding; modes and operating durations as given in Table 3.*

The science goals require the delivery of D-InSAR, polarimetry, and stereo data at Reconnaissance (30 m resolution), Exploration (6 m resolution) and Locality (1 m resolution) scales, as well as radiometry measurements at the Global scale. Note that InSAR is here used to refer to stripmap swaths optimised for repeat-pass D-InSAR; in the strict sense, the first pass acquires SAR, the second InSAR, and the third D-InSAR.

VenSAR will acquire these data in five modes: InSAR (VI1 as standard, VI2 for orbit-to-orbit, and VI3 for opposite-look), stereo polarimetry (VP1 StereoPolSAR), all at Reconnaissance scale (30 m resolution); Exploration scale imagery (VH1 HiRes at 6 m resolution); Locality-scale Sliding Spotlight (VS1 Spotlight at 1 m resolution); and microwave brightness temperature (VR1 Radiometry), as summarised in Table 3 and illustrated in Figure 9.

*Table 3   Summary of VenSAR Operating Mode Parameters*

| | Resolution | Looks | Tx | Incidence | Sensitivity | Swath | Duration | Data |
|---|---|---|---|---|---|---|---|---|
| VI1 InSAR | 27 m | 18 | 4% | 21° – 31° | −21·8 dB | 53 km | 498 s | 66 Mbps |
| VI2 InSAR | 27 m | 18 | 4% | 19° – 29° | −20·9 dB | 53 km | 498 s | 68 Mbps |
| VI3 InSAR | 27 m | 18 | 4% | −21° – −31° | −21·8 dB | 53 km | 498 s | 66 Mbps |
| VP1 StereoPolSAR | 30 m | 9 | 4% | 37° – 41° | −16·9 dB | 53 km | 873 s | 127 Mbps |
| VH1 HiRes | 6 m | 6 | 20% | 38° – 43° | −20·1 dB | 22 km | 291 s | 353 Mbps |
| VH2 HiRes | 6 m | 6 | 20% | 38° – 43° | −20·1 dB | 32 km | 291 s | 513 Mbps |
| VS1 Spotlight | 1 m | 1 | 20% | 38° – 39° | −21·5 dB | 5 km | 4 s | 468 Mbps |
| VR1 Radiometry | 5 × 30 km | n/a | 0% | −4° – +4° | ~1 K | 38 km | <2760 s | <0·25 kbps |

**Figure 9    VenSAR Mode Sensitivities versus Backscatter**

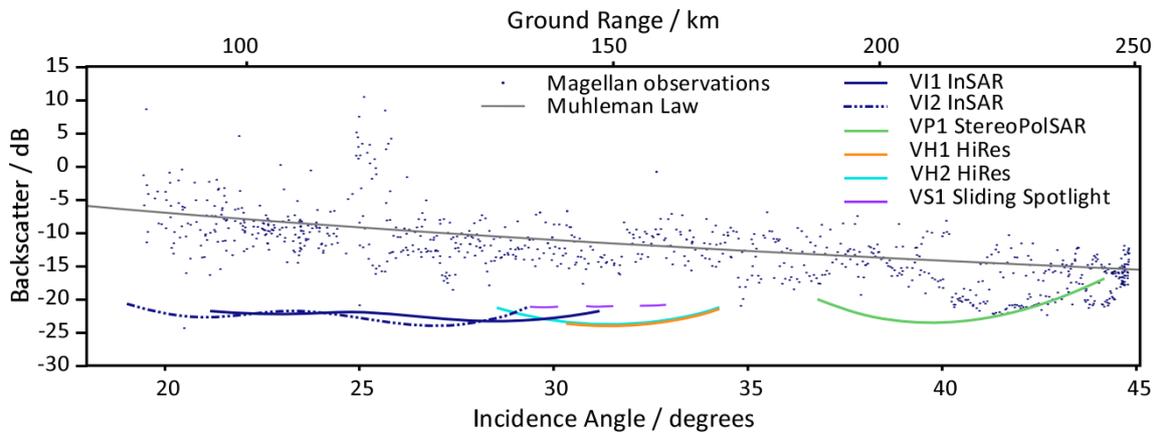

It is, however, possible to programme VenSAR for any other desired mode, incidence angle, or resolution, at any stage of the mission, making it a highly responsive system. However, for nominal mission plan uses repeated daily blocs, with the same pre-defined operation modes to observe different targets, in order to minimise operations complexity and therefore cost. For each mode the radar antenna will be physically pointed towards the optimum illumination angle for each swath by rotation of the whole spacecraft about its roll axis; electronic beam control is used to optimise ambiguity performance.

### 3.1.2.   27 m Interferometry Stripmap

For a short period between Cycles 1 and 2 (i.e. between the first and second sidereal rotation periods of 243 days), Magellan was instructed to extend the radar burst duration across the North Pole of Venus to test for the viability of obtaining interferometric data. The results demonstrate that the atmosphere of Venus is stable over periods of at least 7½ hours.

EnVision will acquire two sets of interferometric SAR data, VI1 and VI2, in Stripmap (a continuously imaged swath) mode, with the second stripmap ~90 minutes after the first, on the following orbit. During the delay between these two passes, Venus' slow rotation will cause the ground tracks of the two passes to be displaced from each other by 10 km at the equator.

**Figure 10    Magellan interferogram of the Venus North Pole**

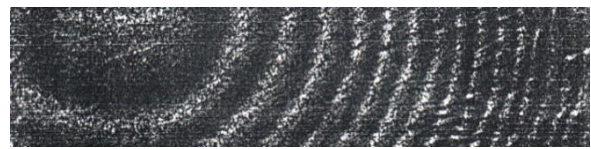

*Goldstein, pers. comm.*

Electronic steering available in the NovaSAR radar technology will be used to repoint the radar beam for the second pass. This baseline would be too large to maintain coherence between the images using a common carrier frequency, but shifts in the carrier frequency between the two acquisitions enable the two data sets to be brought back into coherence[55,102]. The required frequency shifts are on the order of 150 MHz and lie comfortably within the operating spectrum of the radar technology. The long spatial baseline increases the ratio of the topographic phase signal to atmospheric artefacts and other noise, improving the vertical resolution of the topographic model (DEM) produced. Orbit to orbit interferometry ensures that this baseline DEM is obtained within the first Cycle of the mission.

From the second Cycle onwards, EnVision will acquire VI1 and VI3 InSAR on consecutive orbits. VI1 will provide left-looking, and VI3 right-looking, repeat pass (Cycle-to-Cycle) coverage so that the east-west and vertical components of ground displacement may be resolved by comparing the line-of-sight changes in each D-InSAR stack[97]. It is not possible to resolve the north-south component from a polar orbit but it can be inferred from the geological context[59].

**Figure 11   Identifying Ground Deformation with D-InSAR**

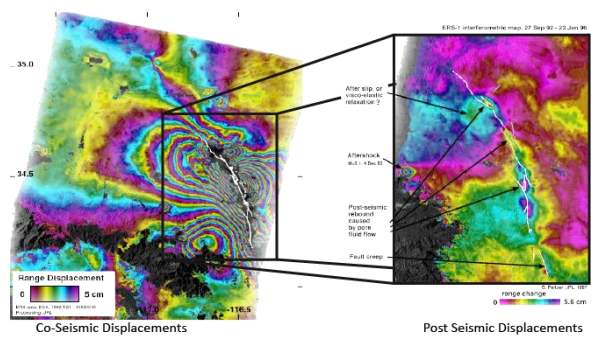

*Ground deformation can be detected with D-InSAR whether or not an earthquake takes place. This example from the 1992 Landers earthquake in California shows the earthquake displacements observed with InSAR (left) and the changes that occurred after the earthquake itself[115] (right).*

In addition to geological change detection, InSAR data are required to measure precisely the precession of the polar axis (Figure 10) and for monitoring the variability in the spin rate. During each InSAR orbit, At the poles the image swaths cross over, leading to a small baseline and allowing many swaths to be stacked coherently. EnVision takes advantage of this to provide a frequent measure of the orientation of the polar axis to monitor changes, with VI1 imaging the North Pole and VI3 the South Pole. The axial precession rate may be resolvable from these data over the lifetime of the mission. To resolve variability in the spin rate and for geodetic control, EnVision will image a contiguous set of swaths across the equator of Venus, connected to the known landing probe locations (discussed later). From the precise timing measurements and subpixel alignment accuracy between InSAR pairs, and using the lander locations as geodetic control points, variability in the spin rate can be monitored throughout the mission.

### 3.1.3. *30 m Stereo Polarimetric ScanSAR*

VenSAR's polarimetric ScanSAR mode (Figure 12) transmits alternating bursts of horizontal and vertical polarisations, with its single receive channel receiving either H or V polarised echoes. Combinations of these options allows a mix of HH, VH, HV and VV polarised images to be obtained. However, this burst mode of operation causes gain variations (image scalloping) and also degrades the image resolution by a factor of NM + 1, where N is the number of polarisation states, and M, the number of looks taken to mitigate scalloping, typically 2. To reduce the total data volume, only one of the two cross-polarised images will be acquired so that the degradation is a factor of 7, enabling a spatial resolution of 30 m.

The InSAR incidence angle (Table 3) is chosen for the optimum phase quality; for polarimetry a higher incidence angle is favoured for its greater sensitivity to surface texture rather than slope. Given this, an angular separation of ~20° has been chosen to allow for the derivation of topography from stereo pairs.

Topography from stereo and InSAR are complementary, in the sense that the former better in steep, rough topography and the latter better in smooth, gently undulating areas. Both approaches provide for a vertical resolution of ~15 m at a spatial resolution of 90~120 m. The resolution, swath width and coverage of InSAR and StereoPolSAR data are purposefully compatible to enable provision of contiguous swaths of interferometric, polarimetric and topographic data across 1500 × 1500 km areas for Reconnaissance mapping.

**Figure 12       Simulated VenSAR Imagery**

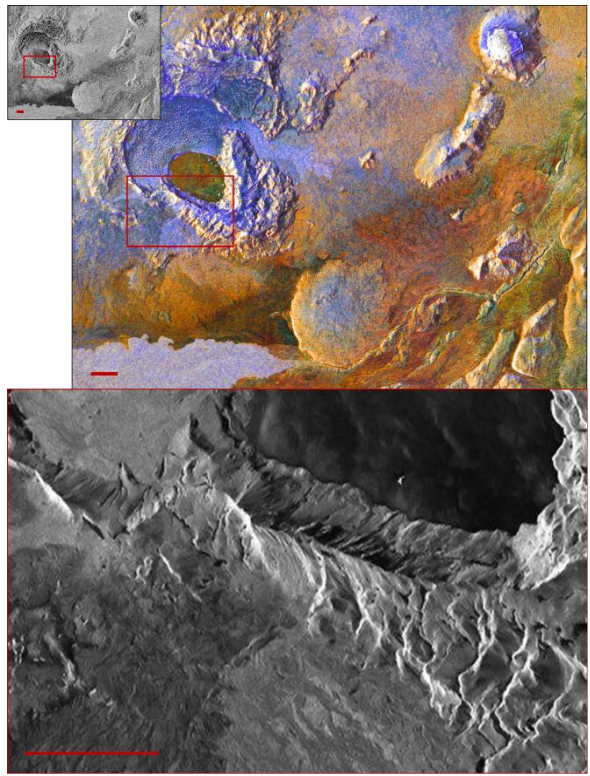

*Simulated VenSAR image products from Holuhraun, Iceland. Top Left: Simulated Magellan 110 m resolution SAR image (derived from Sentinel 1a data). Notice low contrast from 2-bit BAQ compression and foreshortening due to lack of appropriate DEM. Upper Right: Simulated 30 m resolution HHVHVV StereoPolSAR image (derived from Sentinel 1a data). Note the new lava flow in blue at lower left. Bottom: Simulated 6 m resolution HiRes image (derived from TerraSAR-X data). Scale bar in all images is 2 km.*

### 3.1.4.  *6 m High Resolution Stripmap*

Exploration mapping requires 6 m resolution images across selected 100 × 100 km areas. This resolution is achieved in Stripmap mode by increasing the transmit duty ratio (Tx) to bandwidth to 130 MHz – still well within the operating margins – to provide a range resolution of 2 m, which with an azimuth resolution of 3 m, provides for an acceptable 6 looks (Magellan typically had either 5 or 6 looks).

A particular goal for Exploration mapping is the detection of the various lander probes on the Venus surface. The radar cross section of the 2-m diameter Venera landers is approximately 5 dB m², giving a normalised radar cross section (NRCS) >20 dB brighter than the background plains, brighter than any natural feature on Venus and readily distinguishable in single look (2 × 3 m) high resolution data. Once located, even higher resolution Locality imaging will be used to confirm their location and characterise the landing sites.

### 3.1.5.  *1 m Sliding Spotlight*

Having identified the brightest single spot within the landing circle, VenSAR will use Sliding Spotlight to image the landing area at 1 m resolution, with less distortion than Staring Spotlight. In Sliding Spotlight, the radar beam is electronically focussed across a single 5 × 5 km area, instead of the normal continuous Stripmap or ScanSAR methods. By adjusting the incidence angle, three or more Sliding Spotlight images may be taken of the same area, allowing for different polarisation states and stereo pairs to fully characterise the site at the metre-scale.

Up to five 5 × 5 km Spotlight scenes may be acquired in any InSAR or StereoPolSAR orbit, on the opposite node, as indicated in Figure 8. Nearly 450 000 km² of allowing for the imaging of many hundreds of different geological features at the Locality scale during the mission, fully meeting the science requirements.

### 3.1.6.  *Passive Radiometry*

VenSAR's receive and other circuits will remain live throughout each orbit when not actively transmitting (active mode imaging). When the antenna is physically pointed towards nadir for VEM observations, VenSAR will record the brightness temperature of the Venus surface at 3·2 GHz (passive mode), with a precision of ~1 K and a resolution of 4·5 km in azimuth and 38 km in range, a significant improvement on Magellan data. However, without additional equipment the absolute accuracy is only ~160 K due to an uncalibrated radar backend gain. However because the surface temperature is extremely uniform at a given altitude and the variability in backend gain is expected to be small (a few K per orbit), the data can be corrected to provide high quality maps of relative emissivity. To improve the absolute accuracy, the receive circuits can be alternately switched between a known (internal) source and the Venus surface. Integrating the surface signal for 50 ms leads to calibrated absolute brightness temperature accuracy of ~15 K, measured at 2 K precision, at a spatial resolution of 9 km in azimuth and 38 km in range.

### 3.1.7. *Data Processing and Products*

The raw SAR data acquired in all active modes will be losslessly compressed using an optimised block adaptive quantisation method (FD-BAQ), as used on Sentinel-1, reducing the raw data volume by two-thirds, to the values given in Table 3, for their storage and transmission to Earth.

A key aim for EnVision is to have these compressed raw data returned to Earth as Level-0 products, containing the compressed and unprocessed instrument source packets, with additional annotations and auxiliary information to support further processing. These data will be archived for permanent access but not generally released.

Level-1 products are maintained for public access through a web-based map interface and are the normal raw data product format from which all higher level products can be derived. For each acquisition mode, focused Single Look Complex (SLC) products and Ground Range Detected (GRD) products will be generated. All Level-1 products will be georeferenced and time tagged with zero Doppler time at the centre of the swath. Maintenance of these data is vital for research purposes and to allow for future improvements in processing capability and techniques. We anticipate developing components within ESA's Sentinels Application Platform (SNAP) software to allow users to easily and freely acquire and process VenSAR data.

In addition to distributing the Level 1 SLC data, the team will produce and distribute a number of Level-2 and Level-3 processed data types to attract the widest possible audience for EnVision data, including schools and interested members of the public. Level-2 products will include orbit–orbit interferogram products and interim geocoded image mosaics (including multi-polarimetric). Level-3 data products will include ground surface change and deformation maps, absolute Digital Elevation Models of selected regions, gridded at 60 m, and final image mosaics, processed and orthorectified, as single images and multi-polarimetric image composites.

Delivery of these various products will follow the timetable indicated in Table 4. All data products will be compliant with Planetary Data System (PDS4) standards, with product file and folder naming conventions paralleling those adopted for ESA Sentinel-1 data products, and will include descriptors for mission, sensor, mode/beam, product type, processing level, product class, polarization, start and stop dates and times, orbit number, plus a unique product identifier. Files names will also include standard file extensions to indicate file format.

Stereo and InSAR image pairs can be quickly processed to derive relative elevation models, but final image mosaicking and regional gridding to absolute elevation data will be an iterative process which will involve improving the geodetic solution using the sub-surface sounder and gravity data, and refinement of the orbital ephemeris data.

Beyond the data used by the EnVision team for generating DEMs and conducting their own ground deformation and change-detection science investigations, there will be thousands of potential InSAR scenes and stereo pairs for future investigators to use.

A searchable graphical database portal will be maintained to enable investigators to find suitable SLC data for their own processing, or search and download higher level data products. File conversion routines to convert VenSAR data into compatible formats for common non-proprietary SAR, InSAR and radargrammetry software packages will also be provided.

*Table 4        VenSAR Product Delivery*

*Level 1 data, delivered within 6 months of data collection, at 3 month intervals*

- Single-look complex images (SLC)
- Multi-look map oriented images (GRD)
- Radiometry profiles
- Initial ephemeris data

*Level 2 data, delivered within 6 months of the completion of each cycle*

- Interim geocoded image mosaics (single & multi-polarimetric)
- Orbit-orbit interferograms
- Revised ephemeris data

*Level 3 data, delivered within 1 year of completion of prime mission*

- Ground surface change and deformation maps
- Final image mosaics, geocoded & orthorectified
- Digital Elevation Models (gridded at 60 m)

### 3.2.    Subsurface Sounder

The use of a low frequency nadir looking radar sounder provides the ideal complementary information to the SAR data acquired by the S-band VenSAR, enabling a full and detailed investigation of the surface and subsurface geology of Venus. A radar sounder operating at VHF or UHF central frequencies can acquire fundamental information on subsurface geology by mapping the vertical structure (mechanical and dielectric interfaces) and properties of tesserae, plains, lava flows and impact debris. It also provides information on the surface in terms of roughness, composition and permittivity (dielectric) properties at wavelengths completely different from those of SAR, thus allowing a better understanding of the surface properties. The combination of InSAR data (intensity, topography and displacement variables) with the sounder data results in an exceptional ability to understand the link between the surface and subsurface processes on Venus.

SRS is a nadir-looking radar sounder instrument which transmits low frequency radio waves with the unique capability to penetrate into the subsurface. When these radio waves travel through the subsurface, their reflected signal varies through interaction with subsurface horizons and structures with differing dielectric constants. These varying reflections are detected by the radar sounder and used to create a depth image of the subsurface (referred to as radargram) and so map unexposed subsurface features. The design of the SRS instrument depends on the physical and electromagnetic modelling of the surface and subsurface targets. Magellan radar measurements of surface dielectric properties and roughness distinguish two major terrain types:

- Highland areas which are mostly characterised by high values of the dielectric permittivity $\varepsilon$ (>20) and high surface roughness that may limit the sounder penetration.
- Smoother lowland areas with $\varepsilon = 4 \cdot 8 \pm 0 \cdot 9$ and probably high porosity that are suitable for sounding, covering ~80% of the surface.

The resolution and depth of penetration of the sounder depend on frequency. The maximum plasma frequency on the day side is 5-6 MHz, and below 1 MHz on the night side; signals below those frequencies cannot propagate to the surface. The sounder's radar signal will be distorted as it crosses the dispersive plasma of the ionosphere but this distortion can be corrected. The correction algorithm provides information on the total electron content of the ionosphere, which is important for calibrating the phase information recorded by VenSAR.

The sounder returns also suffer from clutter (Figure 13) caused by off-nadir surface reflections reaching the radar at the same time as subsurface nadir reflections and potentially masking them[45]. The strength of clutter is controlled by statistical parameters of the topography that scatters the radiation, which can be derived from the stereo data acquired by VenSAR.

**Figure 13** **Example of clutter removal from subsurface radar data.**

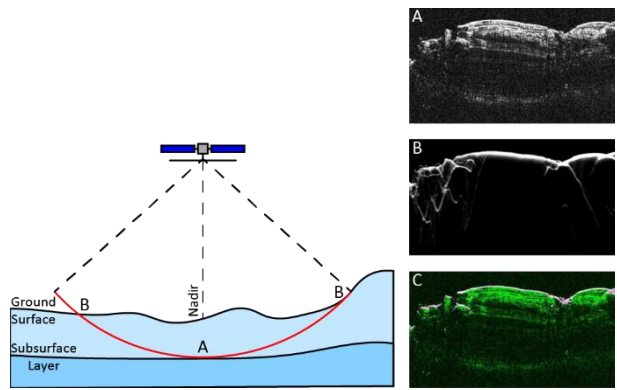

Studies and experience gained with MARSIS and SHARAD on areas of Mars with similar properties to those expected at Venus demonstrate the feasibility of subsurface sounding to provide a complete picture of near subsurface properties. As Venus has higher temperatures than Mars, it is important to point out that the dielectric properties of rocks and, in particular the loss tangent which is the parameter controlling the attenuation, depend on polarization and conduction phenomena. The polarization term is only slightly affected by temperature, whereas conductivity is strongly affected by such parameter. However, in the MHz range the latter term is not predominant. This has been confirmed by dielectric measurements on both Moon samples and terrestrial basaltic rocks[21]. Thus the relatively high temperature of Venus does not affect the penetration capability of the sounder.

### 3.2.1. *Choice of Central Frequency*

To achieve the science requirements, the radar shall be designed to work with a central frequency in the range 9 to 30 MHz for optimal ground penetration capability. The radar bandwidth shall be of several MHz to achieve adequate range resolution. The SRS maximum penetration depth, which has been inferred from the various dielectric measurements in different types of basaltic rocks[21], is shown in Figure 14. Each curve represents the mean value of different conditions for each scenario. Additional simulation studies confirm these data. The maximum penetration depends on both the central frequency and the subsurface composition. For example, assuming a mean relative permittivity $\varepsilon = 6$ and the mid-case scenario, radar operations at a central frequency of 9 MHz result in an average penetration of 600 m and a range resolution of ~16 m. Higher central frequencies, such as 30 MHz, result in a shallower penetration (~350 m) and an improved resolution of 5 m.

**Figure 14** **SRS Penetration Depth**

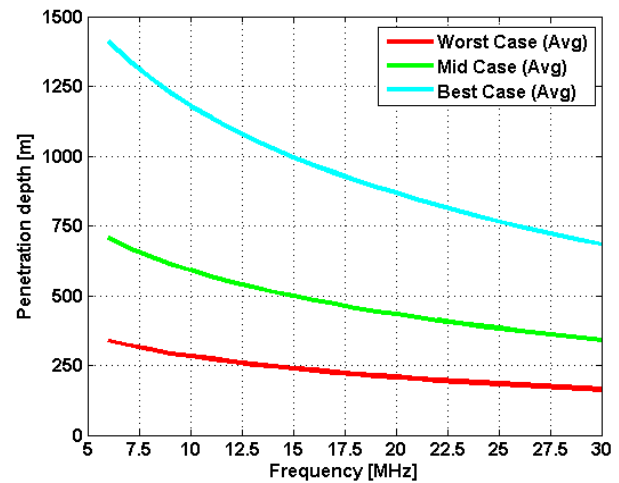

The value of the range resolution for each central frequency and for a varying terrain dielectric constant is shown in Figure 15 where the radar bandwidth is assumed to be equal to 67% of the respective central frequency. In the design phase there will the opportunity to trade range resolution for penetration depth and vice versa.

High Resolution/High Penetration and Very High Resolution/Shallow Penetration are both interesting configurations in terms of science return but for the nominal mission a central frequency of 16 MHz is adopted. However, a full analysis and final decision will be made during the Phase A study.

**Figure 15    Range Resolution versus Frequency**

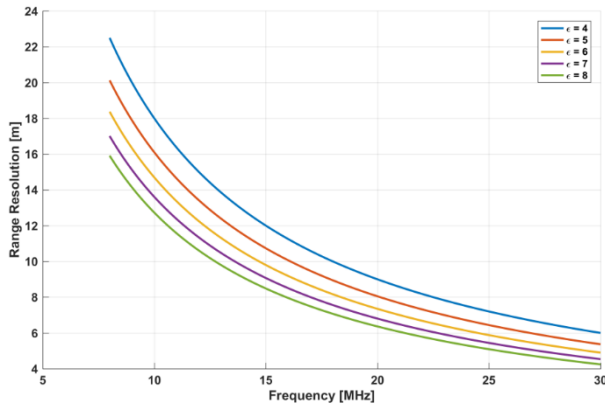

*Table 5    SRS Main Instrument Parameters*

| Parameter | Possible Range |
|---|---|
| Central Frequency (fc) | 9 to 30 MHz |
| Bandwidth | 6 to 20 MHz |
| Antenna Length | ≤16 m |
| Maximum Depth | <1200 m |
| Range Resolution | <6000 m |
| Azimuth Resolution | <1700 m |
| Vertical Resolution ($\varepsilon = 6$) | ~16 m to ~5 m |
| Raw Data Rate | 4.5 to 15 Mbits $s^{-1}$ |
| Compressed Data Rate | 1 to 3.7 Mbits $s^{-1}$ |
| Mean Power Consumption | 50 W |
| Instrument Mass | 10 kg |

### 3.2.2. Sounder Design

SRS consists of a deployable dipole antenna and an instrument block (Figure 16) consisting of two main parts: the receiver and digital subsystem (RDS), and the transmitter and matching network. The instrument is divided into two main parts: the receiver and digital subsystem (RDS), containing the receiver module and the digital electronics section (DES) that includes the digital and conversion functions; and the transmitter (Tx) and matching network, which provide the high power amplification for the signal for transmission and impedance matching to the antenna.

**Figure 16    SRS instrument block diagram**

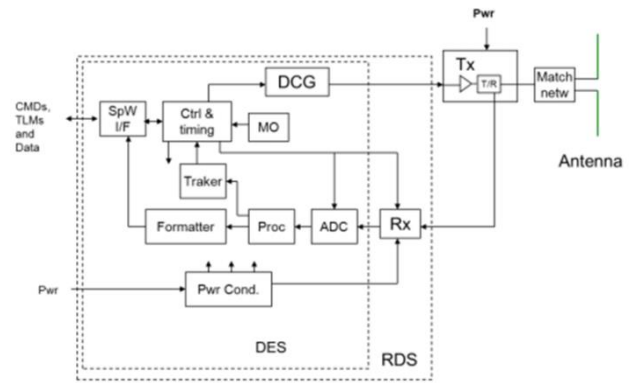

The SRS data processing flow and the resulting data products are presented in Figure 17. The pipeline for Level 1b (L1b) processing will use as input both scientific and engineering telemetry of the instrument, and the orbital information of the spacecraft. Data will be split into individual files, one for each type of telemetry and for every observation, with packets arranged in time order, and corrupted or duplicated packets removed. For scientific telemetry packets, a set of parameters describing the geometry of observation will be computed using orbital information of the spacecraft.

**Figure 17    Scheme illustrating data processing flow and related data products**

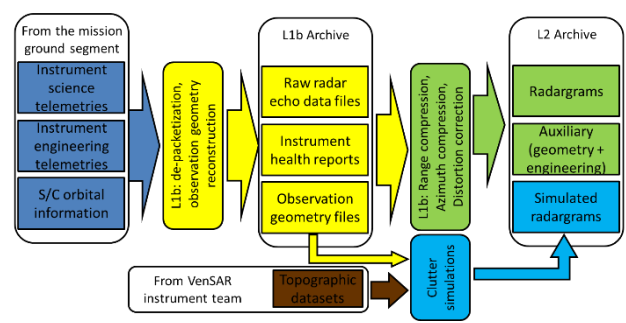

The main steps in L2 processing are azimuth compression (trough SAR processing), range compression and, if needed, correction of ionosphere distortion. Azimuth compression will consist of the sum of a number of echoes, after compensation for their relative delay as the distance of the spacecraft from the target changes along the trajectory. If plasma is present along the line of propagation of the pulse, it will act as a dispersive medium, resulting in the defocusing of the range-processed data. This effect is present in MARSIS data, and several algorithms have been devised to correct it. The product of L2 processing is a set of echoes expressed in units related to voltage measured in the receiver, which can be analysed together as a radargram.

The simulation of surface echo by using DEMs is very important for clutter reduction[45], so that this a critical point of the data analysis. For SRS, the simulation will be established as part of the data processing ground segment: the simulated data will be delivered, archived and disseminated as auxiliary data of the radar profiles that will be included in the L2 data archive. Some geophysical parameters, such as the dielectric permittivity of the surface, can be derived by co-processing the simulated and real data.

SRS heritage includes RIME (Radar for Icy Moon Exploration) onboard JUICE (Jupiter Icy Moon Explorer), MARSIS (Mars Express) and SHARAD on NASA's Mars Reconnaissance Orbiter.

### 3.3. *Venus Emission Mapper*

The Venus Emission Mapper (VEM) instrument suite consists of three channels: VEM-M, VEM-H and VEM-U. VEM-M will provide near-global compositional data on rock types, weathering, and crustal evolution by mapping the Venus surface in five atmospheric windows. The broadest window at 1·02 μm is mapped with two filters to obtain information on the shape of the window. Additional filters are used to remove clouds, water, and stray light. VEM-M will use the methodology pioneered by VIRTIS on Venus Express but with more and wider spectral bands, the VenSAR-derived DEM, and EnVision's circular orbit to deliver near-global multichannel spectroscopy with wider spectral coverage and an order of magnitude improvement in sensitivity.

VEM-H will be dedicated to extremely high-resolution atmospheric measurements. The main objective of the VEM-H instrument is to detect and quantify $SO_2$, $H_2O$ and HDO in the lower atmosphere, to enable characterisation of volcanic plumes and other sources of gas exchange with the surface of Venus, complementing VenSAR and VEM-M surface and SRS subsurface observations. A nadir pointed high-resolution infrared spectrometer is the ideal instrument for these observations at the 1·0 μm, 1·7 μm, and 2·0 – 2·3 μm atmospheric windows that permit measurements of the lower atmosphere. Baseline observations will be performed on the night side but observations at all times of day are possible.

VEM-U will monitor sulphured minor species (mainly SO and $SO_2$) and the as yet unknown UV absorber in Venusian upper clouds and just above. It will therefore complement the two other channels by investigating how the upper atmosphere interacts with the lower atmosphere, and especially characterise to which extent outgassing processes such as volcanic plumes are able to disturb the atmosphere through the thick Venusian clouds. A moderate-resolution (~500 to 1000 cm$^{-1}$) spectral imager in the 200-400 nm range able to operate both in nadir and stellar occultation range would be especially suited to such a task. VEM-M will obtain repeated imagery of surface thermal emission, constraining current rates of volcanic activity following earlier observations from Venus Express.

VEM-M and VEM-H channels will use a MERTIS-derived instrument controller and power supply within a common housing but the different optical requirements need separate apertures and optics to maximise performance with only a marginal mass penalty. In combination, VEM will provide unprecedented insights into the current state of Venus and its past evolution. VEM will perform a comprehensive search for volcanic activity by targeting atmospheric signatures, thermal signatures and compositional signatures, as well as a global map of surface composition.

#### 3.3.1. *VEM-M*

VEM is a pushbroom multispectral imaging system (Figure 18) drawing strongly on DLR's MERTIS instrument for BepiColombo. VEM

incorporates lessons learned from VIRTIS with band centre and width scatter approximately five times more stable than VIRTIS and incorporating a baffle to significantly reduce scattered light and improve sensitivity. The telecentric optics images the scene onto the filter array and relayed by a four-lens objective onto the detector. The filter array is used to provide greater wavelength stability than a grating design. VEM's low development risk results from a standard camera optical design, a flight-proven InGaAs detector with a thermo-electric cooler, and flight-qualified support systems from MERTIS.

**Figure 18    VEM-M design concept.**

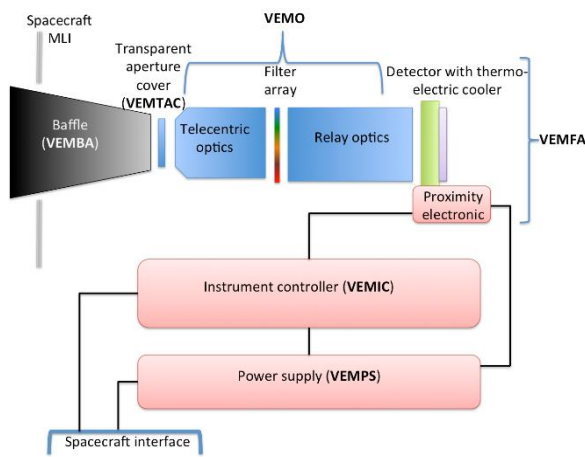

The instantaneous field of view (FOV) of the optics is 45°, equivalent to 307 km from the nominal orbit altitude. Each pixel resolves $0.07° \times 0.07°$ (303 m), which with an integration time of 90 ms, leads to pixel dimension of 303 m across and 1000 m along track. Each ultra-narrow-band filter (made by Materion) occupies 33 of the 640 pixels across track; these are binned along and across track into $10 \times 10$ km cells at the top of the clouds (for cloud correction) and $60 \times 60$ km cells at the surface, providing a SNR of at least 300 for the cloud correction band at <1·5 μm and >500 for the mineralogical bands. This approach provides contiguous spectral emissivity coverage with the 10 km orbit advance.

VEM obtains continuous night-side nadir observations in all spectral bands. To disentangle the surface and atmospheric contributions to the observed radiances, VEM uses an improved version of the extensively tested data pipeline developed to process VIRTIS surface data[107] (Figure 19). The pipeline performs radiometric and geometric corrections and projections, corrects the radiance of surface windows for cloud opacity, and retrieves surface emissivity using topography obtained in parallel by VenSAR. DLR developed the photometric, geometric, and atmospheric corrections for the VIRTIS surface data pipeline and plan to integrate radiative transfer methods[147] during the Phase A study.

**Figure 19    Laboratory model of VEM-M under testing at DLR.**

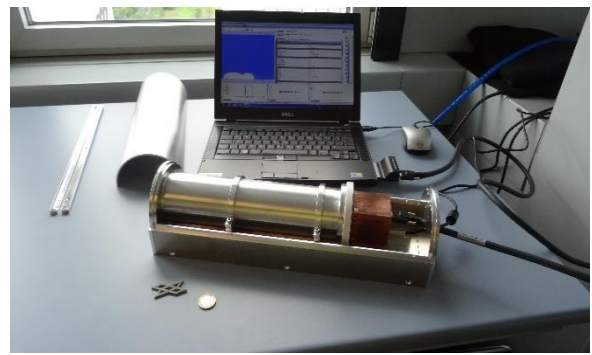

### 3.3.2. *VEM-H*

VEM-H is based on NOMAD (Nadir and Occultation for MArs Discovery), a suite of three spectrometers scheduled to launch on ExoMars Trace Gas Orbiter in January 2016. NOMAD itself has heritage from SOIR (Solar Occultation in the Infrared), which operated on ESA's Venus Express mission. Specifically, VEM-H is a redesign of the LNO (Limb, Nadir and Occultation) channel of NOMAD, retaining much heritage from the original but with minor modifications to meet the science objectives of the EnVision mission.

Like NOMAD, VEM-H is an echelle grating spectrometer coupled to a high-performance, actively-cooled SOFRADIR HgCdTe detector, which utilises an Acousto-Optic Tunable Filter (AOTF) for order selection. These components are optimised for Venus atmospheric observations by shifting the spectral range to 1·0–2·5 μm whilst retaining the TRL and heritage of NOMAD and SOIR. An excellent SNR is achieved through small mechanical modifications: the baseplate is inverted, so that the optical components are directly fixed to the instrument's external-pointing face, which is passively-cooled by a cryo-radiator. The VEM-H electronics are self-contained, allowing mounting on the deck of the spacecraft itself, avoiding additional thermal input into the instrument.

- The optics of the VEM-H channel (Figure 20) are divided into three main units:
- Entrance optics (entrance diameter of 20 mm) that collects the light, defines the FOV and restricts the observed wavelength domain using an AOTF;
- Spectrometer with an echelle grating that defines the free spectral range and the instrument line profile (ILP = 0·2 cm$^{-1}$ FWHM); and
- Detector system that records the spectra according to a spectral sampling interval.

To compact the design, the spectrometer is used in a quasi-Littrow configuration. The collimating and imaging lenses are merged in one off-axis parabolic mirror.

**Figure 20    VEM-H design and photograph of the SOFRADIR detector.**

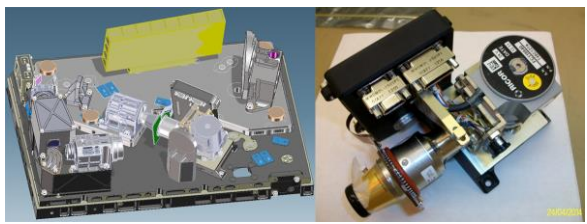

*Left*: the optical assembly, which is inverted and mounted on a baseplate on the underside of the radiator. *Right*: the SOFRADIR detector for NOMAD on ExoMars.

The fast response of the AOTF allows quasi-simultaneous measurements of interesting atmospheric constituents to be performed, through almost instantaneous access into any wavelength domain within the AOTF's frequency parameters. The echelle grating will be manufactured by Advanced Mechanical and Optical Systems (AMOS) Ltd in Belgium. It is a diamond-machined aluminium grating with a blaze angle of 63·2° and a tilt angle of 2·6°. The dimensions of the grating are 150 × 60 × 25 mm and will cover all wavelength orders between 1·0 and 2·5 µm.

The detector is a slightly modified standard Integrated Detector Dewar Cooler Assembly (IDDCA, type ID MM0067) from SOFRADIR, France (see Figure 20). It contains a high-sensitivity Focal Plane Array (FPA) of 30 × 30 µm HgCdTe photovoltaic cells arranged in 320 columns (spectral direction) and 256 rows (spatial direction). The alloy mixing ratio is optimised for the 1·0 and 2·5 µm spectral range of the science requirements at an FPA temperature of 90 K. It is mounted in an evacuated Dewar with a customised optical window and is surrounded by a cold shield with an $f$/4 aperture and cooled by a K508 closed-cycle miniature Stirling cooling machine from RICOR (Israel), adapted for space applications. This configuration has a high heritage from SOIR, NOMAD and many other space missions. The acceptable non-operational temperature range is 243 to 323 K and the SNR at the radiator cooled operating temperature of 240 K is ~230 at 1·17 µm and ~52 at 2·46 µm for a 4 s integration time.

### 3.3.3. *VEM-U*

VEM-U will be based on the heritage from various space-borne UV spectrometers such as SPICAM[14] on Mars Express, SPICAV[15] on Venus Express and PHEBUS[31] on Bepi Colombo. The necessity of observing both faint, point-like sources during stellar occultations and the bright, extended source that Venus cloud top is will be a major driver for the optical design.

In order to fulfil these requirements, the optical scheme will be follow closely the SPICAM-UV/SPICAV-UV philosophy (Figure 21), i.e. a slit of variable width will allow for a trade-off between spectral resolution and radiometric flux, in conjunction with UV gratings for the spectral dispersion. A tunable intensifier will be placed close to the CCD detector. Such a proven design will warrant high performance and flexibility at a very high TRL.

**Figure 21    SPICAV-UV optical scheme.**

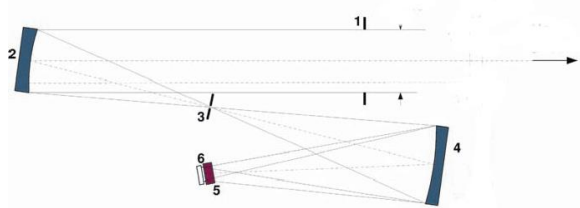

(*1*) Aperture of the UV channel. (*2*) Off-axis parabolic mirror. (*3*) Slit (actuator-controlled wide to narrow). (*4*) Concave UV grating. (*5*) Intensifier. (*6*) CCD.

Based on Venus Express (mostly SPICAV and VIRTIS) observations in both nadir and occultation modes, the design should meet the following specifications:

- Spectral range: 170 – 300 nm; 300 – 450 nm
- Spectral resolution: 0·3 nm @ 170 – 300 nm; 1·5 nm @ 300 – 450 nm

The higher spectral resolution in the 170 – 300 nm interval compared to SPICAV is necessary to retrieve separately the vertical profiles (in occultation) or column densities (in nadir) of SO and $SO_2$ separately with as little degeneracy as possible. Also, extending the spectral range towards the visible spectrum compared to SPICAV will also allow for a complete spectroscopic characterization of the UV absorber whose nature has eluded the scientific community since the 1970s – this is especially important since it could be coupled with the sulphur cycle and possibly with the planet interior through volatile outgassing.

The primary mirror will consist in an off-axis parabola coated with $MgF_2$ and with a focal length around 120 mm. The dispersive elements would be two concave, holographic and toroidal gratings (low-resolution at longer wavelengths, high-resolution at shorter wavelengths) also coated with $MgF_2$. As for the intensifier and detector, we could use similar hardware as on SPICAV: a Hamamatsu intensifier, with a solar blind CsTe photocathode and a $CaF_2 + MgF_2$ input window, followed by a Thomson CCD cooled around 270 K.

### 3.4. *Radio Science*

While not strictly a scientific instrument, EnVision's telemetry system will be used for sounding the neutral atmosphere and ionosphere of Venus, during the frequent occultations that occur during the communications links on the towards inferior and superior conjunctions. As the spacecraft starts to be occulted (or after, during egress) the spacecraft carrier signal probes layers of the planet's atmosphere, causing changes in the frequency and amplitude of the carrier waves (at X- and Ka-bands). The bending that occurs through atmospheric refraction can be retrieved from the Doppler shift residual obtained during the occultation event, with accurate estimates of the spacecraft state vectors[46]. By determining the dependence of the signal bending angle with respect to altitude (more precisely the impact parameter), profiles of the neutral and plasma densities can be derived, essential for characterising the atmospheric structure and dynamics[113,118,144].

By probing the neutral atmosphere above ~ 40 km, density, temperature and pressure profiles can be derived to characterise the atmosphere and its longitudinal and latitudinal distribution. In the same way, by probing the ionosphere above ~ 80 km, its structure can be characterised and its interaction with the solar wind plasma studied. For these experiments, an axi-symmetric model of the atmosphere will be implemented in the data processing in order to better characterise the variability of the neutral atmosphere and ionosphere with latitude.

The experiment is a by-product of the radio tracking required for orbit determination when the spacecraft is occulted by Venus, since it needs no extra instrumentation, only the communications subsystems and VenSAR's ultra-stable clock. Any orbiting spacecraft is sensitive to the local gravity field, plus the gravity field of the Sun and, to a minor extent, other planets. These perturbations, which may be as low as 0·1~0·01 mm s$^{-1}$ for harmonic degrees up to 150, are described by the Lagrange planetary equations and can be solved by the Precise Orbit Determination method[84] using an iterative least-squares fitting of an orbit model to Doppler line of sight (LoS) velocity perturbations and ranging tracking data over successive data-arcs of a few days long, including orbital crossover points. Navigational X-band Doppler tracking data have a precision of 0·02 mm s$^{-1}$ over a 60 second time-count, which allows for the detection of velocity perturbations corresponding to gravity anomalies at a spatial resolution as fine as 125 km. Preliminary simulations indicate that an accuracy of ~±0·002 in $k_2$ is achievable by stacking together 3 years of navigation tracking data of EnVision spacecraft, more than sufficient to distinguish between different models of internal structure. Nevertheless, more realistic simulations are required in order to assess the effect of the gravity field and other perturbing forces (like attitude manoeuvres), as well as improvements possible by combining navigational X-band and science telemetry Ka-band tracking data, on the accuracy of the $k_2$ Love number that can be derived from EnVision.

When conducting this experiment, coherent dual-frequency transmission (X- and Ka-band) is desirable in order to separate the non-dispersive from the dispersive media effects, both to distinguish the neutral atmosphere from the ionosphere, and also to reduce the propagation noise of the tracking observables from solar corona and solar wind effects. The experiment will be conducted in two/three-way mode, as required for navigation purposes so that normally only ingress profiles will result from the experiment. However, the near-polar circular orbit favours the spatial coverage of the profiles sampled. ESTRACK ground stations will provide closed-loop tracking data during the occultation experiment. In addition, the PRIDE technique[41] will be used to simultaneously track the spacecraft with VLBI network radio telescopes and provide open-loop Doppler observables and VLBI observables during the occultation event. The PRIDE open-loop Doppler observables will be used for multipath corrections, essential when characterising the Venus tropopause, and probing the atmosphere at lower altitudes where defocusing and absorption greatly affect the closed-loop Doppler tracking.

In addition to their scientific value, used in combination with Sounder measurements, these data are useful for correcting ionospheric and mesospheric phase delays in VenSAR data and may be obtained as a by-product of the requirement to track and reconstruct EnVision's orbit for accurate InSAR, geodesy and spin rate measurements.

### 4. *Minimum Mission Configuration and Profile*

The nominal launch date for EnVision from Kourou is 24 October 2029, on an Ariane 6.2, with a cruise of 23 weeks, arriving on 5 April 2030. Following capture, apoapse will be lowered to ~50 000 km altitude and periapse lowered progressively to 150~130 km to allow for a ~200-day period of aerobraking to lower the apoapse to ~260 km altitude, after which periapse will be raised to circularise the orbit in November 2030. A walk down phase is required to ensure that the desired dynamic pressure of 0·3 Pa is obtained. Following a period of systems and scientific instrument tests, the nominal mission is scheduled to start on 8 November 2030 and ends on 5 November 2034, having completed 6 Mapping Cycles. Fuel provision is made for ~1500 orbit adjustments (once per day) to maintain the orbit within a 100 m corridor during the Mapping Cycles, required for InSAR. The mission could end as late as 28 February 2035, providing two months of contingency for mapping start delays.

*Table 6    Fuel Requirements*

| Manoeuvre | δV | Fuel† |
|---|---|---|
| Venus Capture | 1050 m s$^{-1}$ | 446 kg |
| Apoapse lowering | 269 m s$^{-1}$ | 95 kg |
| Aerobraking | 120 m s$^{-1}$ | 40 kg |
| In orbit corrections | 127 m s$^{-1}$ | 38 kg |
| **Total** | **1323 m s$^{-1}$** | **619 kg** |

† including 2% margins; Isp = 321 s

The exact trade between initial apoapse lowering and aerobraking duration depends on the Ariane 6.2 launcher capability and hence available wet mass after Venus capture (which is assumed from the M5 Call Annex to be ~1250 kg), prior to apoapse lowering. Should the launcher prove more capable, an additional 100 kg of fuel would reduce the aerobraking duration by ~95 days.

An alternative Venus delivery using solar electric propulsion[6] was studied for EnVision by the Advanced Space Concepts Laboratory at Strathclyde (Appendix B); it could deliver a greater mass into the required <300 km circular orbit but would extend the time required to achieve that orbit by at least 3 years and add significantly to the cost of the mission. Hence our nominal mission design for EnVision uses conventional bipropellant (MON/MMH) fuel and an array of eight 22 N thrusters for Venus capture and orbit manoeuvres.

#### 4.1. *Orbit Selection*

Both interferometric SAR and gravity field measurement require a well−controlled circular orbit, maintained within a 100 m corridor, i.e. no more than ±150 m in x and y, and ±50 m in z (altitude). The resolution of the gravity field that can be measured by the mission declines rapidly with altitude, so that the selected orbit should be as low as possible, which also reduces the fuel required to raise the periapse at the end of aerobraking, and the corrections required for solar perturbations on the orbit. Both Magellan and Venus Express encountered sensible atmosphere below 200 km altitude, and densities sufficient for aerobraking at 130 to 160 km altitude, but the atmospheric density structure is variable and uncertain. Therefore an altitude above ~230 km is desirable.

A study by the Instituto Superior Técnico, Lisbon (Appendix C), optimised the final orbit for the imaging constraints of obtaining InSAR imaging across the equator and North Pole, and repeated high resolution images of the landing sites of Veneras 8, 9, 10, 13 and 14 and Vegas 1 and 2, as early as possible to act as fixed ground control point markers for precise geodesy and to reduce the mission duration to the minimum possible. The optimum orbit obtained is 92 minutes at 259 km altitude, much lower than a typical terrestrial SAR satellite, at 88·2° inclination and with the ascending node at 285·8° longitude. This orbit enables all the landing sites to be imaged within the first 65 days of each cycle, with repeat imaging on the opposite (descending) node ~120 days later, providing for a tight geodetic constraint for precise orbit control.

#### 4.2. *Spacecraft Configuration*

The nominal spacecraft layout (Figure 22) is based on the M4 configuration developed by Thales Alenia Space (Appendix D). Following the pattern of ExoMars TGO, the spacecraft comprises a central structural tube, 937 mm in diameter, supporting a 2 m sided cube. The only deployable systems are the Solar Arrays and SRS; all other scientific instruments are fixed in their final flight configuration, greatly reducing complexity and cost. In the Ariane 6.2 launch vehicle the structural tube is vertical, connected directly at the base (in orbit, the −x direction) to the launch adapter and supporting the 3 m diameter HGA at the top (+x).

The instrument face (in orbit, the nadir, −z, direction) supports the 5·47 m long VenSAR and the 9·4 m long SRS antennas, aligned vertically in the launcher, as well as the VEM telescope aperture, all pointed to nadir. Inboard are the NIA and SRS processors and VEM instrumentation, and the 50 TB solid state memory modules.

Perpendicular to this face, in the −y and +y directions, are the solar arrays, stowed for launch. The −y face is also the cold face containing the radiators, recessed to avoid stray light. The radiators are sized at 30 cm² per dissipated W (indicated in Figure 22), requiring heat pipes for the largest radiators to dissipate heat properly. The VenSAR front end dissipates heat directly from its panels (see Section 3.1.1 and Figure 8); the largest radiators are for VEM and the communications subsystem.

**Figure 22    Nominal Spacecraft Layout**

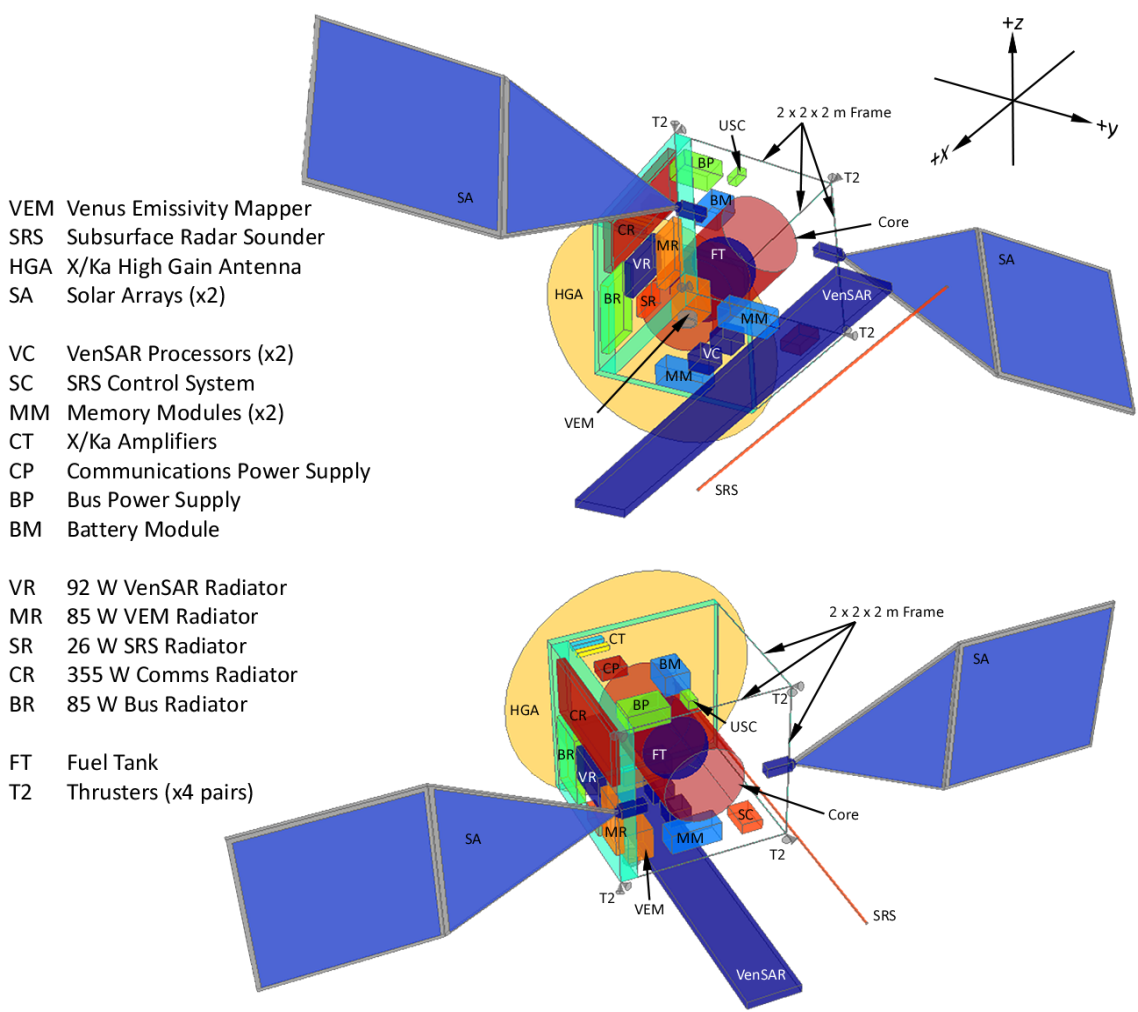

The +z anti-nadir face is free but internally supports the primary bus power distribution subsystem and batteries, and the communications subsystem connected to the +x face HGA. The corners of the -x face support eight 22 N thrusters that comprise the primary propulsion system; no main engine is required. The bi-propellant MON/MMH fuel tanks are housed inside the central structural core. The -x launcher interface face is otherwise free and sun-pointed during ballistic cruise and forward-facing (in the orbit direction) during aerobraking and science observations. The current best estimate mass and power budgets are listed in Table 7. Primary electrical power is provided by two solar arrays providing 2400 W at 15° solar incidence. The yokes also support solar panels both to provide a total of 12 m² for power and to enhance aerodynamic drag for efficient aerobraking. The arrays have one degree of freedom: axial rotation. To avoid overheating the arrays are rotated to avoid full incidence, as adopted for BepiColombo. The sizing case is when the Sun is in the orbital plane. The most consuming mode is during communications following from (or leading to) an InSAR swath; combined with platform requirements, 3400 kJ is required over the 92 minutes of an orbit. Assuming 30° incidence angle, 5580 kJ of usable power is produced per orbit. This covers the consumption with more than 50% margin.

The batteries are sized to allow for science observations and communications on both the day and night sides. Conservatively assuming the highest consumption over one orbit without production from the Solar Array, 2·1 kW.h of usable output is required. With ABSL 18650 NL modules this leads to a battery effective capacity of 2·9 kW.h and a best estimated mass of 18 kg, plus 20% of maturity margin.

*Table 7      EnVision Budgets*

|  | TRL | Margin | Mass | Power |
|---|---|---|---|---|
| Bus Structure and Harness |  |  | 185 kg | 120 W |
| Solar Array and Batteries |  |  | 97 kg | (>2400 W at <15° incidence) |
| Power Management and Electronics |  |  | 58 kg |  |
| Propulsion Subsystems |  |  | 70 kg |  |
| Navigation and Sensors (including CMG) |  |  | 54 kg |  |
| Thermal Control |  |  | 64 kg |  |
| Communications (X/Ka HGA, X LGA) | 6 | 15% | 65 kg | 375 W (during telemetry) |
| VenSAR (front and back end) | 6 | 20% | 176 kg | 112 W (1933 W peak) |
| SRS antenna and electronics | 5 | 20% | 22 kg | 15 W (60 W operating) |
| VEM-M, -H, -U | 5 | 25% | 14 kg | 12 W (23 W operating) |
| **Total Dry** |  |  | **805 kg** | **300 W (2076 W peak)** |
| +20% System Margin |  |  | 949 kg |  |
| Launch Adapter |  |  | 105 kg |  |
| MON/MMH Propellant |  |  | 619 kg |  |
| **Total Launch Mass** |  |  | **1673 kg** |  |

### 4.3.   *Communications*

Our telecommunications strategy assumes one daily communications pass with a 34 m ESTRACK station. The duration of Cebreros to Venus Express visibility periods varied between 6 and 13 hours, depending on season; we have therefore assumed a conservative average of 6 hours per downlink pass. The communications system design is based on specifications in the M5 Call Annexe, supplemented with additional directions provided to us at the M4 Q&A for proposers on 27 November 2015. Accordingly, our nominal mission design utilises a fixed 65 W RF, 3-m Ka-band high gain antenna (HGA), with a modest assumed spacecraft pointing requirement no better than Venus Express (i.e. ~0·06° for 6 hours, tracking Earth). The telemetry scheme has been verified by Airbus Defence and Space (Appendix E) and assumes GMSK (0·5) modulation, used on Solar Orbiter, and 1/4 Turbo codes, used with Bepi Colombo, Juice and Solar Orbiter. Other options include OQPSK modulation and 1/6 Turbo coding but the differences in link budget are within 1 dB. Higher bit-rate modulation schemes, e.g. 8PSK, are possible but not adopted in the baseline case. All science data are assumed to be on Ka-band only, with navigation and engineering command data on X-band, with both bands required for accurate tracking. Use of X-band for additional science downlink, is recommended for study during Phase A.

A communications link pass of 6 hours per day is adopted but inevitably this will often be interrupted by occultations of the spacecraft by Venus. These occultation periods have been included in the link budget calculations with an additional 4 minute margin assumed for data reception check and retrieval procedures following each occultation. The Earth-Venus distance varies by nearly 1·5 AU, causing a factor of 32 difference in the telemetry link budget during the 584-day synodic period (Figure 23). In addition, a 30 day communications blackout is assumed across each superior conjunction. These factors require the adoption of a data collection strategy adapted to the available link capacity.

Sounder data are essential for altimetry, which is used alongside tracking data and orbit crossovers to provide accurate orbit control, while D-InSAR requires repeated observations under near-identical viewing geometries. These data must therefore be obtained throughout the synodic period, requiring storage of data during part of the period; reducing the InSAR data rate is therefore critical. Achieving the optimum scientific benefit from the mission requires a trade-off between InSAR resolution and spatial coverage by InSAR, StereoPolSAR and HiRes modes.

There are no timing constraints on StereoPolSAR and HiRes data, which can therefore be collected when there is sufficient link capacity. The data rates for Radiometry and VEM are sufficiently low to allow their collection throughout the synodic period.

**Figure 23    Nominal Mission Profile**

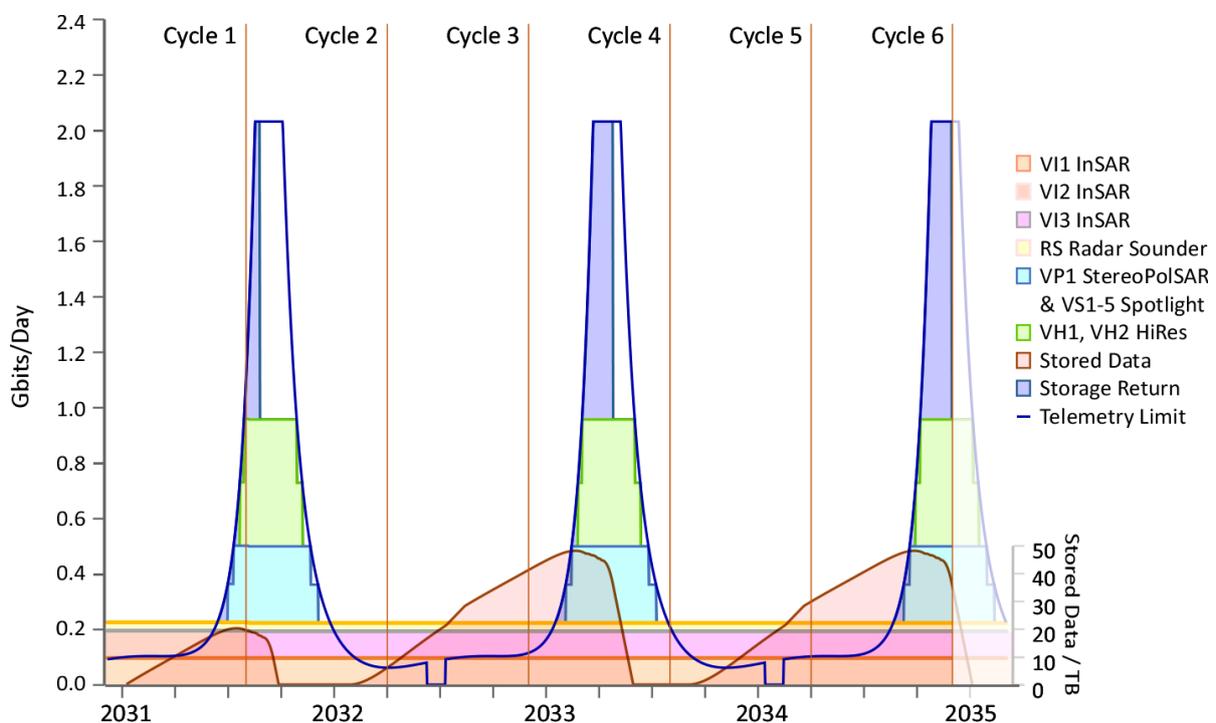

*Imaging strategy is adapted to available telemetry rate: towards superior conjunction only the essential InSAR and sounder data are acquired and stored; as the available bandwidth increases towards inferior conjunction, stored data are returned while StereoPolSAR and HiRes data are also acquired.*

### 4.4. *Scientific Data Collection Strategy*

Once aerobraking is completed, the nominal orbital period is ~92 minutes, which corresponds to 10 km of longitudinal movement of the subnadir point at the equator as the planet slowly rotates under the spacecraft. It takes one Venusian day to sweep through the full range of longitudes. Counting ascending and descending passes, the spacecraft thus passes over every point on the planet twice every Venus day. Whilst normal operations image ahead of the orbit on the ascending node, imaging can occur on the descending node to investigate changes at ~120 days separation, and behind the orbit (opposite-look) on the ascending node, as is the case for VI3, to provide for 1 to 4 days of time separation.

The requirement for contiguous data sets of different types places a constraint of a swath width of at least 40 km in order to span the daily 3-orbit, ~6-hour telemetry link. VenSAR is designed to collect 53-km wide swaths to meet this requirement and the additional 10-km baseline between VI1 and VI2. The subsurface sounder will continuously record data along the nadir track while VEM will operate across the night side of Venus only. Radiometry data are also collected on the night side of every mapping orbit except when VenSAR is actively imaging.

The active imaging strategy depends on the Earth-Venus distance, as noted earlier; InSAR is collected throughout the synodic period but StereoPolSAR and HiRes only when there is sufficient link capacity. In the 24-hour day shown in Figure 24, VI1 is collected during orbits A, F and K and VI2 (or VI3 after Cycle 1) is collected in orbits B, G and N. No other SAR data would be collected at that time. Orbits C, D, and E are dedicated for telecommunication links but by starting after IS2 on orbit B and ending before IS1 on orbit F, a 5·4 hour link duration is obtained.

As the link capacity increases, first VP1 data are acquired in orbits I and then N, and then VH1 data in orbits H and J, and then M and O. During these periods, 5 VS1 Sliding Spotlight images are also obtained on each of the InSAR and StereoPolSAR orbits, except for orbit B. The synodic periodicity of high data rates corresponds to 2·4 Venus days, so that every point on the planet will have had both ascending and descending passes after two high data rate peaks. All portions of the planet are thus accessible for high-resolution and polarimetric imaging during the nominal mission.

VenSAR observations will include both contiguous InSAR observations of an equatorial strip and both poles, and targeted observations of regions of interest (ROI), e.g. as shown in Figure 6. These are sized to the feature of interest but are typically $1500 \times 1500$ km (Table 1), equating to ~25 ROIs (~25% of the surface), sufficient to sample and characterise the variety of terranes on Venus. These ROIs will be imaged in every cycle with IS1 and with IS2 in cycle 1 and IS3 in every cycle thereafter. StereoPolSAR coverage of these same regions will be acquired once during the whole mission, first in the latter part of cycle 1 and the start of cycle 2, with remaining gaps infilled in cycle 4 and cycle 6 (Figure 23). Over the same intervals, more than 1400 HiRes $100 \times 100$ km and 17,500 Spotlight $5 \times 5$ km scenes will also be obtained within each of the ROIs.

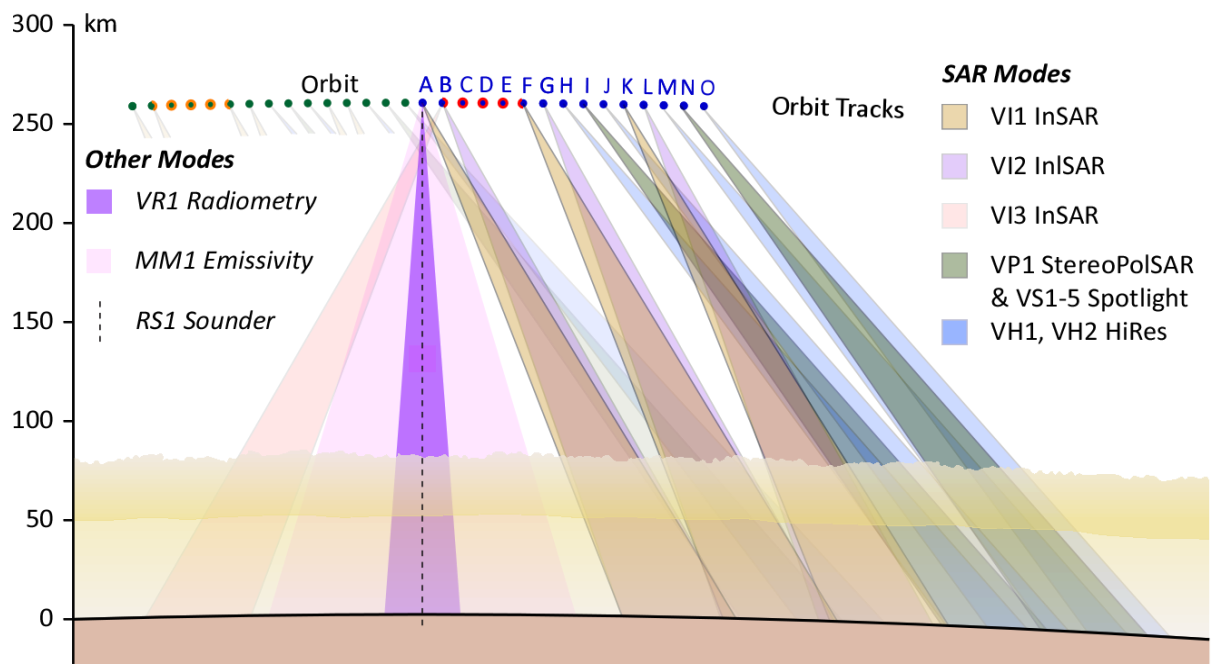

**Figure 24      EnVision Mapping Strategy**

*This rather complex figure illustrates the VenSAR mapping sequence for the ~16 orbits in every 24 hours: 4 orbits are reserved for telemetry (open circles); 3 pairs of orbits for InSAR (VI1 and either VI2 or VI3); 2 orbits for StereoPolSAR; and 4 orbits for HiRes and Spotlight. Sounder, Radiometry and VEM data are collected on every mapping orbit; VEM-M and VEM-H on the night side and VEM-U on the day side.*

Since VEM-M operates on the nocturnal part of EnVision's orbit, coverage is dependent on the interaction between the Venus solar and sidereal days (Figure 25). Complete coverage of the Venus surface takes the full nominal mission of 6 cycles but includes up to 10 repeat passes across most locations, providing ample opportunity for change detection.

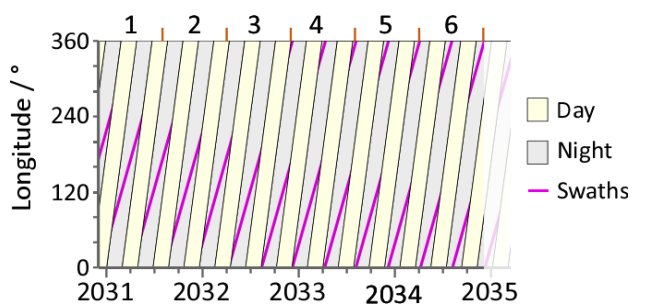

**Figure 25      VEM Mapping Tracks**

## 5. Management Scheme

### 5.1. Procurement

Envision will for the most part follow the procurement approach used in ESA's current suite of planetary missions: the procurement of the spacecraft, launch, mission operations and science operations will be managed and funded by ESA, while the science payloads will primarily be the responsibility of the member states.

VenSAR consists of a front-end antenna of 24 phase centres and back-end electronics for processing and control. Adopting the principle of an integral space telescope, we propose that ESA procure the front-end for integration into the spacecraft structure by the contractor responsible for the spacecraft, and that UKSA provide the backend electronics – effectively the instrument part of the system. Costings assuming this arrangement are discussed below in Section 6.

Likewise, the SRS dipole antenna, including its deployment mechanism, will be procured by ESA, and the backend electronics (the instrument) provided by an Italian-led consortium, funded by ASI.

The VEM instrument will be funded by a German/French/Belgian consortium, with a German PI and French and Belgian Co-PIs. Germany will take responsibility for the instrument and will provide common VEM electronics, management and PA/PP control; France is expected to take responsibility for VEM-U, for the optical components of the VEM-M mapper as well as the procurement of both VEM-M and VEM-H detectors; Belgium will provide the VEM-H spectrometer and its proximity electronics.

### 5.2. Ground Segment

The Ground Segment will follow the model of existing ESA planetary missions: a Mission Operations Centre at ESOC and a Science Operations Centre in ESAC. TM/TC links to spacecraft assume one 35 m ground station for 6 hours per day, for the 5·4 hour communications link. Each science instrument team will be responsible for planning their operations and archiving. Science operations planning follow the ESA planetary mission convention of having a Long Term Plan, defined yearly by the Science Working Team (SWT); Medium Term Plans (MTPs) defined monthly by the SOC working with the science teams; and Short Term Plans (STPs) defined weekly by ESOC for the implementation of telecommands.

The mapping strategy adopted for VenSAR (Section 4.4) not only satisfies the science requirement for delivering nested datasets of target regions, but also simplifies operations planning by using blocks of pre-defined observation sequences in sets consecutive orbits. Operating VenSAR in this way requires spacecraft roll manoeuvres around the x-axis to support different observation modes within each orbit; provision for a control momentum gyroscope (e.g. Airbus CMG 15-45S) has been made for this purpose.

Precise orbit reconstruction for navigation, InSAR, geodesy and spin rate analysis requires regular Doppler tracking of EnVision, preferably during every telemetry link, and preferably at both X- and Ka-bands. These data will complement the altimetric measurements derived from SRS data and location and timing data from VenSAR. Provision of a Galileo-heritage rubidium ultra-stable oscillator (clock) is included for this purpose and to support radio science experiments. These provisions allow for cycle-to-cycle return orbit control to ~100 m; it is anticipated that the reconstructed orbit knowledge will be ~15 m in x, y, and z.

### 5.3. Pointing Requirements and Modes

The pointing needs of the spacecraft are driven mainly by VenSAR, since the antenna must be pointing off-nadir by up to 35° for normal SAR operations. For reliable InSAR, the 3σ requirement is for 30 arcsec (0·15 mrad) over 1000 s. The HGA has a less stringent requirement but over a longer period: 210 arcsec (1·0 mrad) for 21 600 s. These constraints exceed those for VEM (250 arcsec over 2800 s) and SRS, which has the simple observation requirement of operating when the instrument face is nadir-pointing. VEM-M, VEM-H and the

*EnVision* Page 38 of 43

VenSAR radiometer operate only on the night side. VEM can also be operated during night side VenSAR operations when the instrument face is pointed up to 35° from nadir.

The spacecraft shall have at least the following modes:

- **Safe mode**: sun-pointed with launcher face and solar array towards the Sun;
- **Cruise mode**: sun-pointed with launcher face and solar array towards the Sun during ballistic phases;
- **Boost mode**: solar array oriented towards the Sun (one-axis degree of freedom), on day side when no science operations are required;
- **Communications mode**: HGA Earth-pointed, solar array oriented towards the Sun (one-axis degree of freedom)
- **VenSAR mode**: inertial 3-axes pointed, nadir face facing Venus centre, rolled by up to ±35° around the spacecraft velocity vector (x-axis); and
- **Night side science mode**: inertial 3-axes pointed, nadir face facing Venus centre.

The spacecraft shall always be oriented such that the cold face does not receive direct insolation.

### 5.4. *Science Team*

The Science Team organisation is shown in Table 8 below. ESA retains overall control of the mission with an ESA Project Scientist and Mission Manager. There are three science payloads (VenSAR, SRS, VEM) and a Radio Science investigation, each of which is led by a Principal Investigator. The Science Working Team will also include Inter-Disciplinary Scientists and Guest Investigators. The table also shows how responsibility for generation and validation of the different data products is assigned. Working groups on different science themes (e.g. volcanism, tectonics, aeolian features) will be set up to facilitate scientific exploitation between different instrument teams.

### 5.5. *Data Pipeline and Dissemination*

The main data products to be produced from the EnVision mission are outlined in 0. The first stages of the data pipeline (Level 0 – Level 1) will run in an automated process on ESA servers, using algorithms delivered by the instrument teams, so that Level 1 data will be made available within 1 week of data downlink. Level 2 processing will be carried out by instrument teams. All datasets will use data formats compatible with existing software tools, beyond basic PDS 4 compliance. The VenSAR team will deliver software tools to enable reading of VenSAR data with the Sentinel-1 Radar toolkit (part of the SeNtinels Application Platform, SNAP) or equivalent, and with GIS tools such as ENVI.

Topographic and image datasets will also be publicly-accessible via map-based web platforms, as has been done for some Mars datasets, with compressed (JPEG-equivalent) radar images made available in near-real time. Data will be preserved and publicly disseminated using ESA's Planetary Science Archive, in accordance with ESA policy for planetary missions. Each instrument consortium will assign at an early stage individuals responsible for data archiving.

*Table 8     Science Team and Programme Management*

|  | Project Scientist ESA | Mission Manager ESA | Science Archive Co-ordinator ESA | | | | |
|---|---|---|---|---|---|---|---|
| **Instrument** | VenSAR | | | | | Subsurface Radar Sounder (SRS) | |
| **Team Lead (PI)** | **Richard Ghail** Imperial, UK | | | | | **Lorenzo Bruzzone** University of Trento, Italy | |
| **Investigations** | **Stereo SAR** | **D-InSAR** | **HiRes/Spotlight** | **Polarimetry** | **Radiometry** | **Cross-sections** | **Altimetry** |
| **Investigation Lead** | **Robbie Herrick** Fairbanks, US | **Philippa Mason** Imperial, UK | **Richard Ghail** Imperial, UK | **Lynn Carter** Goddard, US | **Alice Le Gall** LATMOS, France | Lorenzo Bruzzone Trento, Italy | Francesca Bovolo FBK, Italy |
| **Level 1 Products** | **V11** Geolocated stereo pair | **V12** SLC data, InSAR DEM, phase | **V13** Hires SAR images | **V14** HH, VH, VV polarised images | **V15** Temperature, S-band emissivity | **S11** Line profile backscatter | **S12** Spot heights |
| **Level 2 Products** | **V21** Geolocated DEMs & SAR | **V22** Coherence & surface motion | **V23** Spotlight SAR images | **V24** Surface scattering | **V25** Permittivity & surface density | **S21** Geolocated cross-sections | **S22** Global topography |

| **Instrument** | Venus Emission Mapper (VEM) | | | Radio Science | | |
|---|---|---|---|---|---|---|
| **Team Lead (PI)** | **Jörn Helbert** DLR, Germany | | | **Caroline Dumoulin & Pascal Rosenblatt** LPG, France & ROB, Belgium | | |
| **Investigations** | **VEM-M** | **VEM-H** | **VEM-U** | **Tracking Data** | **Geodesy** | **Occultation** |
| **Investigation Lead** | **Jörn Helbert** DLR, Germany | **AnnC. Vandaele** BISA, Belgium | **Emmanuel Marcq** LATMOS, France | **Pascal Rosenblatt** ROB, Belgium | **Nicolas Rambaux** IMCCE, France | **Leonid Gurvits** JIVE, Netherlands |
| **Level 1 Products** | **M11** IR emissivity maps | **M12** IR spectra | **M13** UV spectra | **R11** Precise orbit determination | **R12** Geodetic reference frames | **R13** Calibrated Doppler residuals |
| **Level 2 Products** | **M21** Maps of surface characteristics | **M22** Tropospheric trace gases | **M23** Mesospheric trace gases | **R21** Gravity field & geoid | **R22** Spin rate & axis variability | **R23** Atmospheric P,T, $H_2SO_4$ profiles |

### 5.6. *Specific work proposed for Phase A study*

Tasks for Phase A study, apart from the normal tasks of mission design:

- *Additional work on payload definition, including additional hardware for a calibrated VenSAR radiometer.*
- *A development programme to bring the physical structure of the VenSAR antenna to TRL 7.*
- *Improved simulations of the thermal effects on the spacecraft and VenSAR antenna during aerobraking and science phases.*
- *Further work on orbit determination and control requirements for D-InSAR, gravity determination, and geodetic science goals, including a refined calculation of ΔV requirements for orbit control.*
- *Refined simulations of the SRS performance for optimising the choice of radar parameters (e.g. central frequency) with respect to different Venus scenarios (subsurface attenuation, off-nadir clutter, etc.).*
- *Study trade-off between aerobraking duration and fuel capacity.*
- *Detailed ground segment design, in particular for VenSAR (both science operations and data products), to allow a better estimate of science ground segment costs.*

### 5.7. *International Context*

The mission scenario proposed here is an Europe-only mission, with no international participation required. Some NASA/JPL contributions to the SRS instrument are foreseen, following successful previous collaborations on similar instruments, but ASI assumes full responsibility for its funding. In addition to this possible hardware contribution, all three of the science instruments will likely involve NASA-funded science co-investigators, benefiting from the large Venus community existing in the USA. NASA funding of a secondary payload such as a Cupid's Arrow microsatellite (Appendix E) would be another possibility. Participating scientists from Russia, Japan and other nations would also be welcomed.

Planning for ESA Venus missions should also take into account mission proposals under development in other space nations. At the time of writing, two Venus missions are finalists in NASA's Discovery mission category: DAVINCI, an entry probe which would measure atmospheric composition (including noble gas isotopic abundances) and obtain descent imagery of the surface[4]; and VERITAS, an orbiter equipped with an X-band radar[134]. The VERITAS mission's central aim is to provide a global topography dataset at 5 m vertical and 250 m spatial resolution, with standard imagery at 30 m resolution and high resolution imagery at 15 m. It is therefore highly complementary to the EnVision mission, which can obtain imagery at resolutions approaching 1 m, and a higher priority on interferometric ground deformation monitoring (which is more feasible at S-band than at X-band). If the VERITAS mission were selected, it would be launched to Venus in 2022 and conduct a nominal scientific mission from 2022-2025. EnVision would be an ideal follow-up mission to VERITAS: it would benefit from the global topography and gravity datasets obtained from VERITAS, and would be able to follow up with higher resolution imaging, as well as differential InSAR study of target regions identified in the VERITAS dataset. Compared to VERITAS, EnVision will carry an expanded version of VEM with two additional channels VEM-H and VEM-U.

Venus In Situ Exploration is one of the six mission types allowed under NASA's New Frontiers call (proposals due late 2016 for launch in 2024). The in situ measurements from such a mission would be highly complementary to the orbital measurements from EnVision – and more so for the DAVINCI entry probe proposal under consideration in the Discovery mission category, which would image and land in tesserae terrain not previously sampled.

The Russian Federal Space program includes Venera-D, a Venus lander and orbiter. With launch currently foreseen as 2026+, the development of the Russian mission is on a similar timescale to the ESA M4 programme. The Venera-D lander will measure surface mineralogy and lower atmosphere gases, including isotopic abundances of noble gases, to help to constrain the formation and evolution of Venus, while the orbiter focusses on atmospheric and ionospheric measurements. All these investigations complement those of EnVision; collaboration between an Envision-derived orbiter and a Venera-D lander would offer an L-class level of science return. One productive scenario for collaboration would be to have independently launched EnVision and Venera-D missions, with one or more Russian contributed scientific payloads on EnVision and European contributed payloads on Venera-D. Another scheme could involve an EnVision orbiter and Venera-D entry probe on a single large launcher, similar to the ExoMars TGO-EDM. Inclusion of proximity radio links, which would enable the Envision orbiter to provide data relay an entry probe, lander or balloon, is another way to exploit co-operation between the missions. Thirdly, mutual radio occultation between the orbiters could use the existing communications systems of the orbiters to study the upper atmosphere and ionosphere with high vertical resolution and much better coverage than would be achieved with Earth–spacecraft occultations alone.

We note also that Venus missions are (or have been) under study in India, China, and Japan. It is possible that one of these nations will launch a Venus mission within the next decade or two. Such missions would serve to increase our understanding of Venus, without providing the sophistication and scientific power of a Envision's scientific payload.

### 5.8. *Outreach*

Planetary missions offer tremendous opportunities for public outreach and EnVision already has a web (*www.envisionvenus.net*) and social media presence sufficient to attract national media. Some examples of outreach products produced by the European Science community can be seen at *www.eurovenus.eu*.

EnVision will produce terabytes of high-resolution topography and imagery, well-suited for outreach, all publicly available in near-real time. We will investigate possibilities of publicising these using well-known, publicly accessible portals such as Google Earth (following the example set by the near-real time release of MRO/THEMIS data in NASA's "Live from Mars" project). Through our Outreach Coordinator the Envision team is engaged with "Citizen Science" projects (e.g. zooniverse.org) to widen public engagement in the project. We propose that the ESA mission budget should include at least two person-years for an Outreach Coordinator, appointed by and working within ESA's Communications Department, to oversee press releases, media events and education programmes in co-ordination with EnVision science teams. This would cover the period from six months before launch in 2029 to the end of nominal science mission in 2035.

### 5.9. *Opportunities for secondary payloads*

EnVision is a straightforward one-spacecraft mission with three scientific payloads (VenSAR, VEM and SRS). However, our notional design includes an unallocated spacecraft face (+z, opposite the VenSAR antenna) that could be used to accommodate secondary payloads, in the instance that further margin becomes available depending on the final lift capabilities of the Ariane 6.2 launcher. Two examples of secondary cubesat-class subsatellites that would be highly complementary to the scientific goals of EnVision and further increase its impact are outlined in Appendix F. Other examples could include payloads focussing on measuring solar wind interaction and atmospheric escape from a dedicated microsatellite. A call for secondary cubesats, similar to that recently issued from the AIM asteroid mission, would be an opportunity for enhanced outreach and to infuse new scientific ideas ahead of the foreseen 2029 launch date. To provide data relay for these secondary payloads, and/or in situ mission elements that may be launched after EnVision, we recommend that ESA considers the inclusion of a proximity data relay transponder, like the Electra UHF Mars transponder.

## 6. Costing

EnVision will launch on an Ariane 6.2, costing 73 M€. The nominal mission is planned for an October 2029 launch and last 5 years, to the end of Cycle 6 in November 2034. Scaling overall mission length from the written M4 debrief (Appendix G), Mission Operations Costs are assumed to total 65 M€ but because the total data volume is significantly lower, Science Operations Costs and ESA Project Team costs are assumed unchanged at 27 M€ and 51 M€ respectively.

The industrial cost is expected to be less than M4 because the design is simpler, with a fixed X/Ka HGA and only one degree of freedom on the solar arrays, and with only two deployable mechanisms, for SRS and for the solar arrays. The dry mass is consequently reduced; scaling from the M4 debrief leads to 231 M€.

Airbus Defence and Space Ltd estimate that VenSAR, the main payload component, will cost £35·5M, slightly less than at M4 because of the simpler fixed antenna structure. The front end, including the antenna structure, transmit and receive modules, can be regarded as equivalent to an optical telescope, providing for a wide range of different science observations, while the back end is the scientific instrument. Hence we propose that UKSA funds the £9M back end, and ESA funds the £26·5M front end.

Assuming a worst-case exchange rate of 1€ = £0·75, this leads to an ESA contribution of 35 M€ for the front end, effectively amounting to a 5 M€ contingency at the current exchange rate. As noted in Section 3.1, this cost is significantly lower than Sentinel-1 and TerraSAR-X type SARs because of the tenfold technological improvement in phase centre power output, which means that fewer than $1/6^{th}$ the number of phase centres are required (24 compared with ~150). No other savings from mass production, low cost hardware, or other such reductions are assumed. However, an additional 12 M€ has been added to the front end cost to allow for Venus-specific qualification, including further radiation hardening, and the possible addition of separate radiometer equipment. The NovaSAR SAR, from which VenSAR is derived, is being launched into Earth orbit for a total cost to UKSA of £25M, including the spacecraft platform.

SRS is likewise funded by ASI, Italy, for the back end electronics and by ESA for the antenna and deployment mechanism, included in the overall industrial cost, as was the case at M4. VEM is entirely contributed, funded jointly by Germany, France and Belgium.

Given the simplifications to the spacecraft design, communications, and overall mission profile, a contingency of 50 M€ (~10%) is assumed, as recommended in the Call Annex documents, leading to a projected ESA cost at completion of 544 M€.

*Table 9    Projected Costs for EnVision*

| Cost | M€ |
|---:|---:|
| ESA Project Team | 51 |
| Industrial Cost | 231 |
| Payload Contribution (ESA) | 47 |
| Mission Operations (MOC) | 65 |
| Science Operations (SOC) | 27 |
| Launcher | 73 |
| Contingency (10%) | 50 |
| *Total* | *544* |

## Annex A. References

22  Bullock, M. A. & Grinspoon, D. H. The stability of climate on Venus. *Journal of Geophysical Research: Planets* **101**, 7521-7529, doi:10.1029/95JE03862 (1996).

23  Campbell, B. A. Merging Magellan Emissivity and SAR Data for Analysis of Venus Surface Dielectric Properties. *Icarus* **112**, 187-203, doi:http://dx.doi.org/10.1006/icar.1994.1177 (1994).

24  Campbell, B. A. Surface formation rates and impact crater densities on Venus. *Journal of Geophysical Research: Planets* **104**, 21951-21955, doi:10.1029/1998je000607 (1999).

25  Campbell, B. A., Campbell, D. B. & DeVries, C. H. Surface processes in the Venus highlands: Results from analysis of Magellan and Arecibo data. *Journal of Geophysical Research: Planets* **104**, 1897-1916, doi:10.1029/1998je900022 (1999).

26  Campbell, B. A. *et al.* Evidence for crater ejecta on Venus tessera terrain from Earth-based radar images. *Icarus* **250**, 123-130, doi:http://dx.doi.org/10.1016/j.icarus.2014.11.025 (2015).

27  Campbell, M. J. & Ulrichs, J. Electrical properties of rocks and their significance for lunar radar observations. *Journal of Geophysical Research* **74**, 5867-5881, doi:10.1029/JB074i025p05867 (1969).

28  Carter, L. M., Campbell, D. B. & Campbell, B. A. Volcanic deposits in shield fields and highland regions on Venus: Surface properties from radar polarimetry. *Journal of Geophysical Research* **111**, doi:10.1029/2005je002519 (2006).

29  Carter, L. M., Campbell, D. B. & Campbell, B. A. Geologic Studies of Planetary Surfaces Using Radar Polarimetric Imaging. *Proceedings of the IEEE* **99**, 770-782, doi:10.1109/JPROC.2010.2099090 (2011).

30  Carter, L. M., Campbell, D. B., Margot, J. L. & Campbell, B. A. in *37th Lunar and Planetary Science* 2261 (Lunar and Planetary Institute, 2006).

31  Chassefière, E. *et al.* PHEBUS: A double ultraviolet spectrometer to observe Mercury's exosphere. *Planetary and Space Science* **58**, 201-223, doi:http://dx.doi.org/10.1016/j.pss.2008.05.018 (2010).

32  Chetty, T. R. K., Venkatrayudu, M. & Venkatasivappa, V. Structural architecture and a new tectonic perspective of Ovda Regio, Venus. *Planetary and Space Science* **58**, 1286-1297, doi:http://dx.doi.org/10.1016/j.pss.2010.05.010 (2010).

33  Cohen, M. A. B., Lancashire, D. C., Larkins, A., Watson, P. & Lau Semedo, P.    (ESA, ESTEC, Noordwijk, 2014).

34  Cohen, M. A. B., Lau Semedo, P. & Hall, C. D.    (ESA, ESTEC, Noordwijk, 2014).

35  Cottereau, L., Rambaux, N., Lebonnois, S. & Souchay, J. The various contributions in Venus rotation rate and LOD. *Astronomy & Astrophysics* **531**, A45, doi:10.1051/0004-6361/201116606 (2011).

36  Crisp, D., Allen, D. A., Grinspoon, D. H. & Pollack, J. B. The dark side of Venus: near-infrared images and spectra from the Anglo-Australian observatory. *Science* **253**, 1263-1266 (1991).

37  Davies, M. E. *et al.* The rotation period, direction of the North Pole, and geodetic control network of Venus. *Journal of Geophysical Research: Planets* **97**, 13141-13151, doi:10.1029/92JE01166 (1992).

38  de Bergh, C. *et al.* The composition of the atmosphere of Venus below 100km altitude: An overview. *Planetary and Space Science* **54**, 1389-1397, doi:10.1016/j.pss.2006.04.020 (2006).

39  DeShon, H. R., Young, D. A. & Hansen, V. L. Geologic evolution of southern Rusalka Planitia, Venus. *Journal of Geophysical Research: Planets* **105**, 6983-6995, doi:10.1029/1999je001155 (2000).

40  Ding, X.-l., Li, Z.-w., Zhu, J.-j., Feng, G.-c. & Long, J.-p. Atmospheric Effects on InSAR Measurements and Their Mitigation. *Sensors* **8**, doi:10.3390/s8095426 (2008).

41  Duev, D. A. *et al.* Spacecraft VLBI and Doppler tracking: algorithms and implementation. *A&A* **541** (2012).

42  Dumoulin, C., Tobie, G., Verhoeven, O., Rosenblatt, P. & Rambaux, N. in *International Venus Conference*   (ed Colin F. Wilson) 44-45 (Oxford, 2016).
ESA M5 proposal - downloaded from ArXiV.org